\documentclass[%
 reprint,
 superscriptaddress,
 grouped address,
 amsmath,assume,
 aps,
materials,
]{revtex4-2}

\usepackage{graphicx}
\usepackage{dcolumn}
\usepackage{bm}
\usepackage{physics}
\usepackage{xcolor}
\usepackage[normalem]{ulem}
\usepackage{hyperref}
\usepackage{float}
\usepackage{blindtext}
\usepackage{blindtext}

\usepackage{float}

\hypersetup{
     colorlinks   = true,
     citecolor    = blue,
     linkcolor    = blue,
     urlcolor     = blue,
}

\newcommand{\RNum}[1]{\uppercase\expandafter{\romannumeral #1\relax}}

\newcommand{\Rnum}[1]{\lowercase\expandafter{\romannumeral #1\relax}}

\begin{document}

\preprint{APS/123-QED}

\title{A general-purpose atomic cluster expansion interatomic potential for niobium}

\author{Aleksei~Egorov}
\email{aleksei.egorov.research@gmail.com}
\altaffiliation{current address: Center for Advanced Systems Understanding (CASUS), G\"orlitz, 02826, Germany and Helmholtz-Zentrum Dresden-Rossendorf (HZDR), Dresden, 01328, Germany}
\affiliation{ ICAMS, Ruhr-Universit\"{a}t Bochum, Universit\"{a}tstr. 150, 44780 Bochum, Germany }

\author{Ralf~Drautz}
\affiliation{ ICAMS, Ruhr-Universit\"{a}t Bochum, Universit\"{a}tstr. 150, 44780 Bochum, Germany }

\author{Thomas~Hammerschmidt}
\affiliation{ ICAMS, Ruhr-Universit\"{a}t Bochum, Universit\"{a}tstr. 150, 44780 Bochum, Germany }

\begin{abstract}

Niobium, a body-centered cubic transition metal, poses a challenge for interatomic potentials, which struggle to capture its properties, such as phonons, high-pressure behavior, energy barriers to dislocation glide, and others.
To tackle this challenge, we constructed a general-purpose atomic cluster expansion (ACE) potential for niobium. We trained our ACE on thousands of density~functional theory (DFT) structures spanning a diversity of local environments. We validated it across a range of properties and compared it with existing empirical and machine learning (ML) potentials, including a novel universal ML potential. The resulting ACE balances accuracy, efficiency, and robustness, enabling large-scale exploration of niobium with \mbox{near-DFT} precision. Finally, our ACE held its own in a stringent test: a near-million-atom molecular dynamics simulation of fracture.

\end{abstract}

\maketitle

\section{\label{sec:Introduction}Introduction}

Discovered by the English chemist Charles Hatchett in 1801~\cite{nb_discovery_10.1098/rstl.1802.0005}, niobium stands out among body-centered cubic (bcc) transition metals: it exhibits the highest superconducting transition temperature, not only among pure transition metals but across all elements of the periodic table~\cite{highest_Tc_2016electronic}; its intricate electronic structure gives rise to notable anomalies, such as the softening of phonon modes~\cite{Nb_phonon_anomaly_1_PhysRevB.19.6142,finnins_pettifor_Nb_phonon_anomaly_1.5_PhysRevLett.52.291,Nb_phonon_anomaly_3_PhysRevB.82.144114,Nb_phonon_anomaly_4_Liu2011,Nb_phonon_anomaly_5_computation6020029,Nb_phonon_anomaly_6_DFT_pnonon_band_struc_PhysRevB.101.115119}; at high temperatures and pressures, niobium transforms from the cubic bcc phase to an orthorhombic Pnma~\cite{PT_phase_diagram_Errandonea2020}. In niobium, factors like these make it challenging to accurately describe atomic interactions—the mainstay of reliable atomistic simulations.

Quantum mechanical methods---firstly, Density Functional Theory~(DFT)~\cite{PhysRev.136.B864-DFT-1,PhysRev.140.A1133DFT-2}---describe atomic interactions with immaculate precision~\cite{dft_accuracy_Science2017}. But the tardy computing pace (DFT can only deal with a few hundred atoms) often renders them futile for simulations of dislocations, grain boundaries, or cracks, which may demand hundreds of thousands of atoms.
Empirical interatomic potentials, such as the embedded-atom method (EAM)~\cite{EAM}, allow handling large numbers of atoms easily, but they often fail to reproduce DFT or experimental data, rendering any findings dubious~\cite{emp_pot_limit_RODNEY2008418,emp_pots_limit_Beyerlein,moller_bitzek_2014comparative_eam_drawbacks,class_pot_limitations_Bitzek2015,eam_artifacts_HIREMATH2022111283, Rodney_disl_entropy_2025}. 

For a long time, the trade-off between precision and speed barred the way for reliable large-scale atomistic simulations. This quandary was resolved, at least in part, with the advent of interatomic potentials based on machine learning (ML)~\cite{ML_pots_review_drautz_JACOBS2025101214}. If fitted and validated carefully~\cite{morrow_deringer_how_to_validate}, ML potentials provide a confluence of \mbox{DFT-like} precision with speed. Trained on energies and forces from thousands of small-scale DFT calculations, a robust ML potential can replicate them and make predictions beyond. Over the past two decades, a tapestry of ML interatomic potentials has emerged~\cite{NNP_Behler_Parinello_PhysRevLett.98.146401,GAP_PhysRevLett.104.136403,SNAP_THOMPSON2015316,MTP_Shapeev,Drautz-2019-PhysRevB.99.014104,Pickard_EDDP_PhysRevB.106.014102,MACE_NEURIPS2022_4a36c3c5,Kozinsky_potential_Batzner2022,CACE_Cheng2024,MLTB_Burrill2025} and they have held their own across a rich repertoire of \mbox{applications~\cite{menno_bokdam_nobel_PhysRevLett.122.225701,Kormann_Kostiuchenko2019,cheng_water_doi:10.1073/pnas.1815117116,hydrogen_Cheng2020,Deringer_Pickard_boron_PhysRevLett.120.156001,surface_PhysRevLett.125.206101,Oganov_MLIPs_for_CSP_PhysRevB.99.064114,tuckerman_cendagorta2021enhanced,Deringer_Nature_2021,alchemical_PhysRevMaterials.7.045802,nanoconfined_water_PD_Kapil2022,zhang_NPJ_2023atomistic,Meta_Lan2023,Deringer_Device_GAP_Zhou2023,deringer_device_ACE_zhou2025fullcycledevicescalesimulationsmemory,freitas_doi:10.1073/pnas.2322962121,Grabowski_origin_XU2024120423,Beyerlein_mlpot_WANG2021110364,minaam_ngoipala2025hydride,W_diffusion_Grabowski_Zhang2025, Thygesen_2D_van_der_Waals, Pickard_bespoke_utility,rodney2026revisiting_quantum_effects_dislocation}.}
\mbox{And yet,} for large-scale simulations, not all are fast enough or transferable~\cite{MISHIN2021116980,roadmap_Zhang_2025}.

To open the way for fast and reliable atomistic simulations of niobium, we constructed a general-purpose interatomic potential. We leveraged an atomic cluster expansion (ACE)~\cite{Drautz-2019-PhysRevB.99.014104}, which rivals the best \mbox{ML} potentials in accuracy while boasting exceptional \mbox{speed~\cite{pacemaker-Lysogorskiy2021,zhang-maresca-2023GAPs,Leimeroth_Albe_comparison_2025}}. We first expanded an existing DFT reference database for niobium~\cite{GAP_Nb_PhysRevMaterials.4.093802} and then employed it to train our ACE. We then validated our new ACE across a broad swath of material properties, along the way, weighing it against existing niobium potentials, both empirical and ML-based. Lastly, we put our new ACE through a stress test in a large-scale, high-demand molecular dynamics simulation.

\section{\label{sec:ref_data_and_train}Reference data and training}

\begin{figure*}[!htb]
    \centering
    \includegraphics[width=0.97999\textwidth]{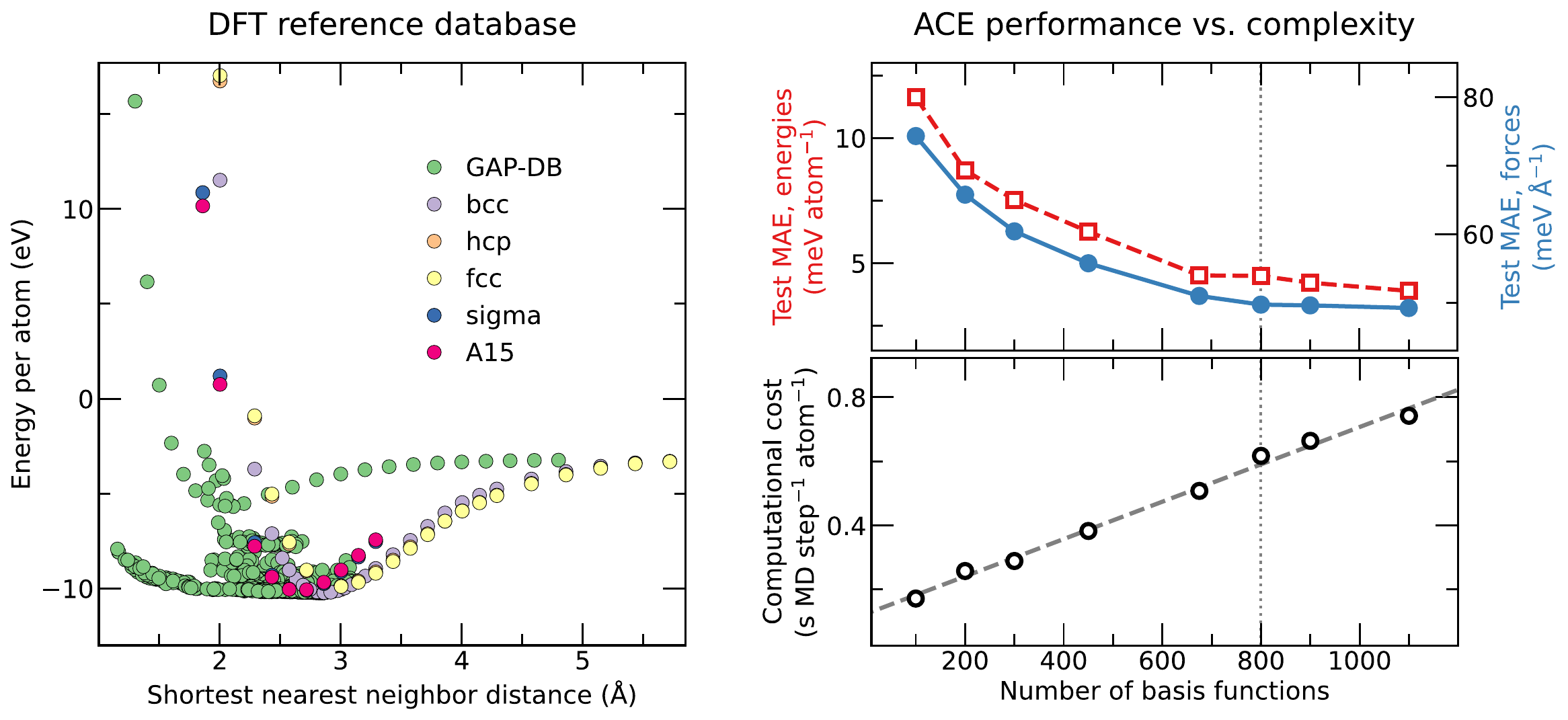}
    \caption{\label{fig:DB_and_perf_vs_bfs} The DFT reference database we employed to train the atomic cluster expansion (ACE) for niobium (left), along with the accuracy and computational cost as functions of the model’s complexity in terms of the number of basis functions (right). The left figure features \mbox{GAP-DB}, the original niobium DFT database by Byggmästar~$et$~$al.$~\cite{GAP_Nb_PhysRevMaterials.4.093802}. It also depicts the structures we added to this database: bcc, hcp, fcc, sigma, and A15. The similarities between fcc and hcp structures and between A15 and sigma result in similar energies; thus, not all data points are visible due to partial overlap. The database was randomly split into 90\% for training and 10\% for testing. The resulting ACE accuracy on the right is depicted as the mean absolute errors (MAE) on the test dataset. The vertical dotted line marks the 800-bf ACE we selected as our final model.}
  \end{figure*}

The quality of a machine learning interatomic potential hinges on the quality of the data for its training~\cite{garbage-in-garbage-out-Deringer2024}. 

To construct a Gaussian Approximation Potential (GAP), Byggmästar~$et$~$al.$ crafted a DFT database for niobium that spans a wide range of properties---bulk, deformed, and rattled bcc structures, point and planar defects, small clusters, and many more~\cite{GAP_Nb_PhysRevMaterials.4.093802}. To train our ACE, we started with their publicly accessible database~\cite{gap_data_nb_training} and expanded it.

We added structures from the DFT energy-volume \mbox{(E-V)} curve of the most stable bcc phase: near-equilibrium volumes to capture the bulk modulus; small volumes to ensure strong repulsion; and large volumes to preclude unphysical behavior such as false local minima.

It is also essential to consider phases beyond the most stable bcc. By their very nature, crystal \mbox{defects exist} out of equilibrium and may form out-of-equilibrium structures \textit{locally}. For example, in bcc metals, the local crystal structure at the screw dislocation cores comes close to that of face-centered cubic (fcc) and hexagonal close-packed (hcp)~\cite{wang2022taming}. For this reason, we enriched the database with both fcc and hcp E-V curves. We further diversified the represented atomic environments with the E-V curves of niobium's second and third most stable phases, A15 and sigma~\cite{Kunzmann_PhysRevMaterials.8.033603}. In their database, the GAP authors included some fcc, hcp, and A15 structures as distorted unit cells near equilibrium volumes~\cite{GAP_Nb_PhysRevMaterials.4.093802}.
\mbox{But to} train the model, which remains sturdy in the face of drastic volumetric upheavals, such as those during fracture, this might be insufficient.
Our final database, depicted in Fig.~\ref{fig:DB_and_perf_vs_bfs} contains 4,130 structures, of which we generated 147 and the rest are from the GAP database. We randomly split it into 90\% for training and 10\% for testing.

In training, the ACE formalism enables increasing the model's complexity by using more basis functions (bf)~\cite{Drautz-2019-PhysRevB.99.014104,Anton-2022-PhysRevMaterials.6.013804}; it helps decrease numerical errors but also raises computational cost. We found that the 800-bf ACE struck the optimum balance and opted for it as our final model (Fig.~\ref{fig:DB_and_perf_vs_bfs}-right). It yields mean \mbox{absolute} errors (MAE) of 4.26 (4.49) meV/atom for energies and 46.91 (49.70) meV/Å for forces in the training (test) dataset. 

\section{\label{sec:validation}Validation and benchmarking}

Minimizing numerical errors alone does not ensure the physically sound behavior of the interatomic potential~\cite{Kovács2021_beyond_rmse,fu2023forcesenoughbenchmarkcritical}; the potential should be \mbox{validated} against key material properties~\cite{morrow_deringer_how_to_validate}. Below, we present the validation of our ACE and compare it with existing interatomic potentials for niobium. For comparison, we considered the original GAP by Byggmästar~$et$~$al.$~\cite{GAP_Nb_PhysRevMaterials.4.093802}, two Moment Tensor Potentials (MTPs)—designated as MTP-1 and MTP-2—from Refs.~\cite{MTP-1_Yin2021} and~\cite{MTP-2_Jung2023}, respectively~\cite{MTP-zotov-comment}. We also evaluated two empirical potentials: the widely used EAM potential~\cite{EAM_PhysRevB.81.144119} and the newly minted extended modified embedded-atom method (XMEAM) potential~\cite{Rui_Wang_phd,xmeam_vanadium_PhysRevMaterials.6.113603}. Given advances in universal interatomic potentials that allow a single model for the entire periodic table~\cite{universal_2_Loew2025,universal_3_Shuang_2025,universal_1_Thygesen_sharma2025acceleratingpointdefectphotoemission}, we considered one at the vanguard of this emerging area~\cite{MatBench_Riebesell2025}: the Graph Atomic Cluster Expansion \mbox{(GRACE)-2L-OAM~\cite{lysogorskiy2025graphatomicclusterexpansion, GRACE_PhysRevX.14.021036}}. Additionally, we evaluated the computational speed of all the potentials.

\subsection{\label{subsec: E-V curves}Energy-volume curves}

\begin{figure*}[!p]
    \centering
    \includegraphics[width=0.945\textwidth]{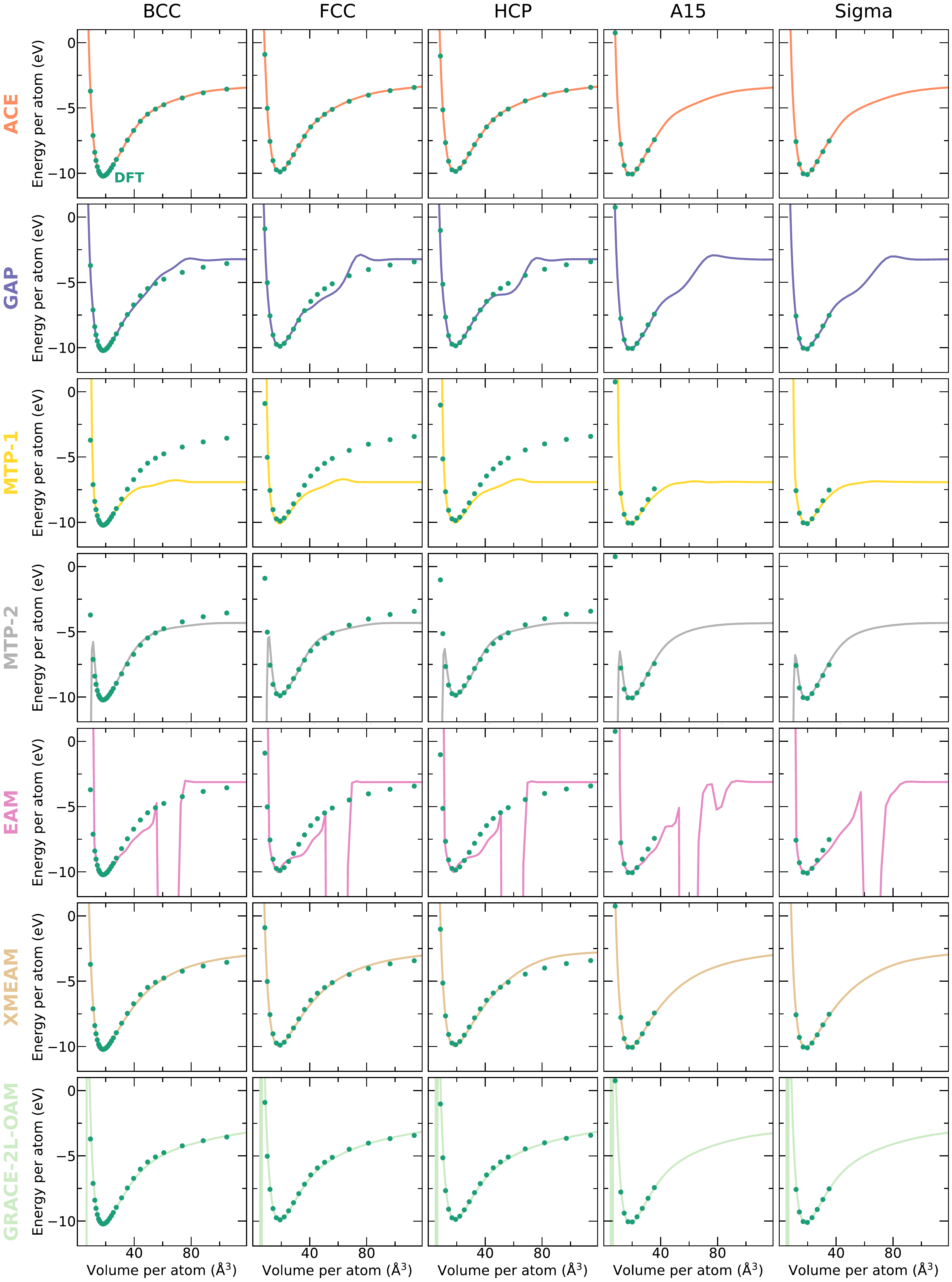}
    \caption{\label{fig:E-V_curves} Energy-volume curves for various crystal structures, obtained with ACE and other interatomic potentials, compared with DFT; the displayed DFT data points were part of the ACE training.
    }
  \end{figure*}

\begin{figure*}[!htb]
\begin{center}
\includegraphics[width=0.589999\textwidth]{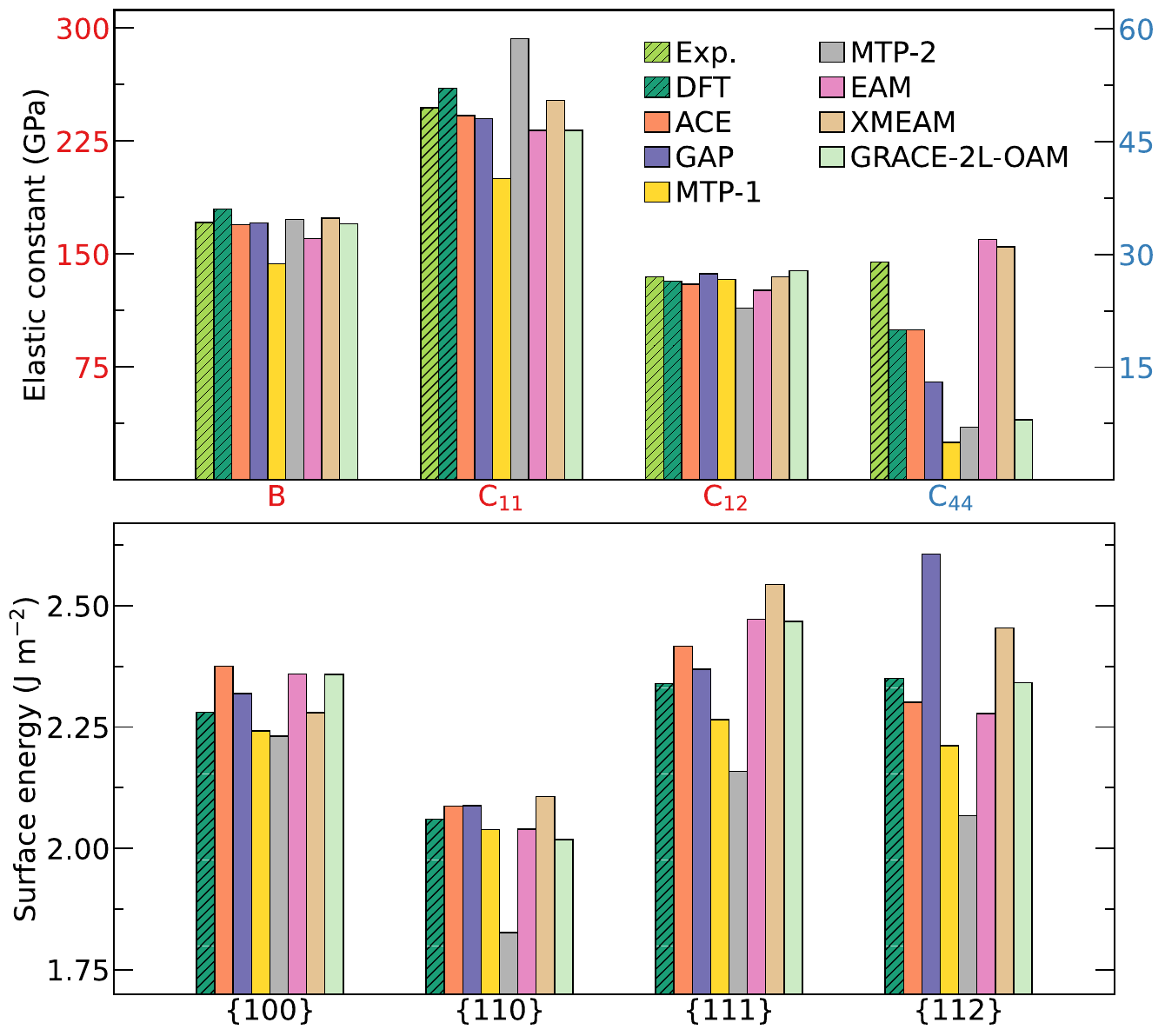}
\caption{Elastic constants (top) and surface energies (bottom) for bcc Nb obtained with ACE and other interatomic potentials, compared with DFT (and, for the former, with experiment~\cite{exp_el_consts_simmons1965single}). DFT surface energies are from Ref.~\cite{surf_ener_materialsproject_Tran2016}.}
\label{fig:el_consts_and_surf_ens}
\end{center}
\end{figure*}

We first computed the energy-volume (E-V) curves for five crystal structures---bcc, fcc, hcp, A15, and sigma (their significance was highlighted in Sec.~\ref{sec:ref_data_and_train}). Our ACE follows the DFT E-V curves across all volumes (Fig.~\ref{fig:E-V_curves}). It smoothly extrapolates to the large volumes of A15 and sigma, despite their absence in the reference data.
The XMEAM, also, by and large, hits the mark.
By contrast, GAP, MTPs, and EAM produce \mbox{E-V} curves that deviate from DFT, especially at large volumes. \mbox{MTP-2} fails to render the repulsion at small volumes, which is bound to cause failures in simulations that involve large departures from equilibrium spacing, such as fracture. Universal GRACE renders accurate \mbox{E-V} curves across all volumes. It as well produces false local minima at small volumes, but only after an immense repulsion barrier; therefore, this is unlikely to cause issues in simulations.

\subsection{\label{subsec: E-V curves}Elastic constants}
Mechanical stability, lattice vibrations, and stress fields around dislocations, crack tips, and other defects—all affected by elastic constants~\cite{el_const_mech_stab_Born_1940,phonons_el_consts_RevModPhys.84.945,anderson2017theory,anderson2005fracture,sutton2024elasticitybook}. We computed the bulk modulus, $B$, and the three independent elastic constants, $C_{11}$, $C_{12}$, and $C_{44}$, for the bcc phase.

ACE results tally with DFT for all elastic constants, reproducing $C_{44}$ exactly (Fig.~\ref{fig:el_consts_and_surf_ens}-top). Yet, it struggles to match the experimental $C_{44}$ because ACE predictions are tied to the DFT training data, and DFT underestimates $C_{44}$ by about 30\%~\cite{dft_c44_nb_underestim_10.1063/1.5136052}. Two empirical potentials, XMEAM and EAM, came close to the experimental value of 29 GPa~\cite{exp_el_consts_simmons1965single}, yielding 31 and 32 GPa, respectively. Meanwhile, ACE, with a $C_{44}$ of 20 GPa, is the closest to the experiment among the ML-based potentials. GAP, MTP-1, MTP-2, and universal GRACE yield, respectively, 13, 5, 7, and 8 GPa.

\subsection{\label{subsec: E-V curves}Surface energies}

Catalysis, fracture, nanoparticle formation, and numerous other processes are rooted in surface energies~\cite{surf_ener_1_JIN2022110029,griffith1921vi,nanopart_surf_ener_relation_Yoko2018}, rendering them one of the key properties a robust interatomic potential must capture. We examined four common bcc surfaces: $\{100\}$, $\{110\}$, $\{111\}$, and $\{112\}$. Their energies, obtained with ACE and other potentials, by and large, aligned with DFT values  (Fig.~\ref{fig:el_consts_and_surf_ens}-bottom).

\subsection{\label{subsec: E-V curves}Traction-separation}

We also examined the so-called traction-separation \mbox{(T-S)} curve for $\{110\}$ planes, which is tied to fracture properties. To produce a T-S curve, a bulk crystal is sliced along a desired plane, and the resulting two halves are rigidly pulled apart. Since DFT cannot handle cracks (due to the large number of atoms required), T-S curves are the closest proxy~\cite{curtin_ts_curve_explained_2005methods,Möller2018_ts_explained}. Additionally, since our training database does not contain T-S curves, evaluating how well ACE \textit{predicts} these \textit{unseen} structures may serve as a test of its transferability.\begin{figure*}[!htb]
    \centering
    \includegraphics[width=0.97999\textwidth]{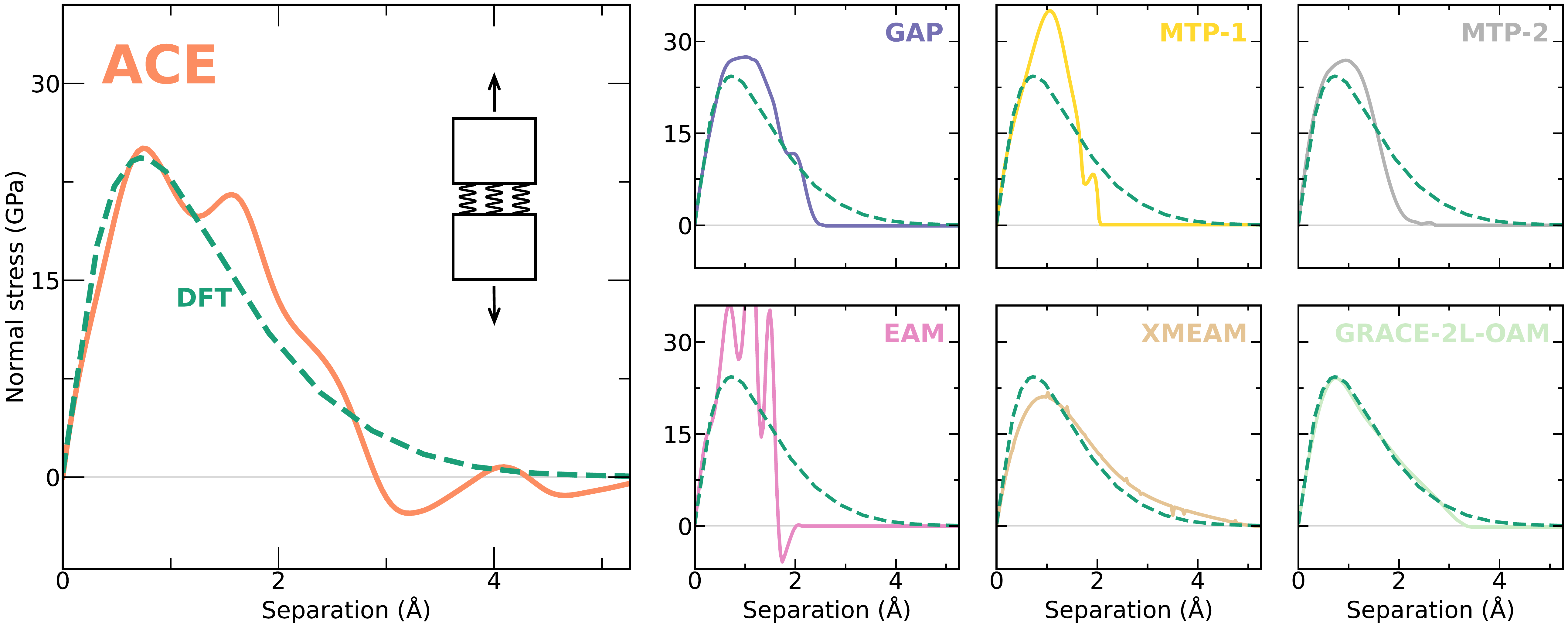}
    \caption{\label{fig:TS} Traction-separation curves for the $\{110\}$ plane of bcc Nb obtained with ACE, other interatomic potentials, and DFT.}
  \end{figure*}

The universal GRACE potential takes the lead, closely following the DFT T-S curve (Fig.~\ref{fig:TS}). For the ACE, key features~\cite{curtin_ts_curve_explained_2005methods}, such as (1) the area under the curve, which delineates the energy dissipated during fracture; (2) the curve's decay length, which defines a zone of nonlinear behavior; and most critically, (3) the peak, which limits the maximum stress the material can withstand, are, by and large, consonant with DFT.

\subsection{\label{subsec: E-V curves}Phonons}

\begin{figure*}[!htb]
\begin{center}
\includegraphics[width=0.97999\textwidth]{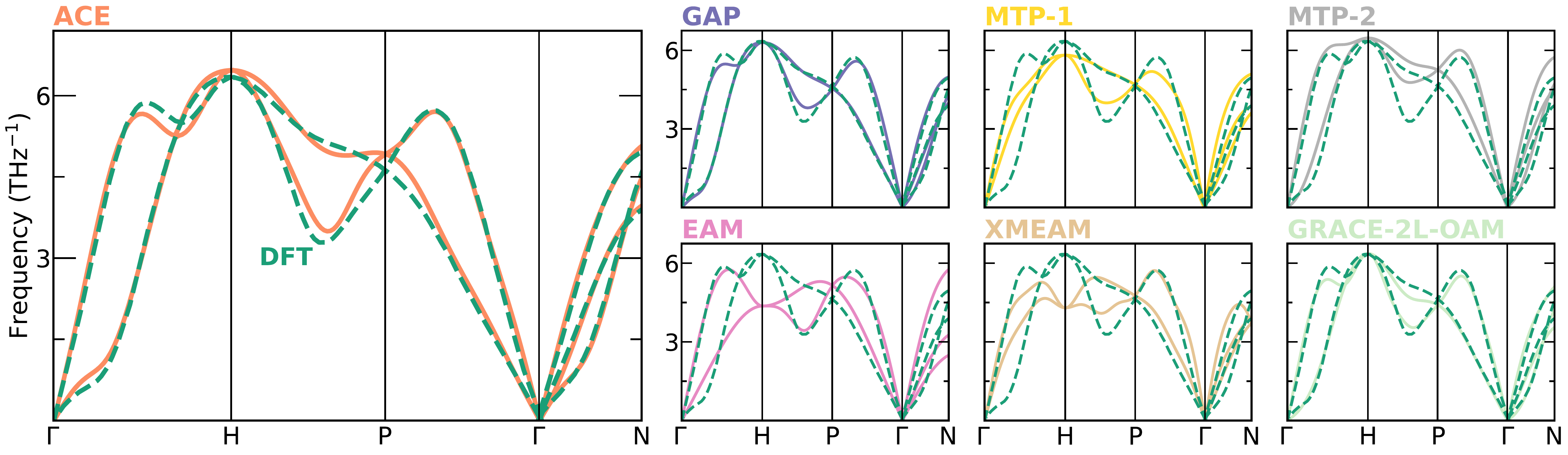}
\caption{Phonon band structure of bcc Nb obtained with ACE and other interatomic potentials, compared with DFT~\cite{Nb_phonon_anomaly_6_DFT_pnonon_band_struc_PhysRevB.101.115119}.
}
\label{fig:phonons}
\end{center}
\end{figure*} For proper behavior of the interatomic potential in finite-temperature simulations, capturing phonons is essential~\cite{pnonons_Jennie_George_Deringer_GAP_10.1063/5.0013826}. Obtaining accurate phonons in bcc niobium is especially challanging due to anomalies in its phonon band structure~\cite{Nb_phonon_anomaly_1_PhysRevB.19.6142,finnins_pettifor_Nb_phonon_anomaly_1.5_PhysRevLett.52.291,Nb_phonon_anomaly_3_PhysRevB.82.144114,Nb_phonon_anomaly_4_Liu2011,Nb_phonon_anomaly_5_computation6020029, Nb_phonon_anomaly_6_DFT_pnonon_band_struc_PhysRevB.101.115119}. The computed band structures revealed that the ACE, GAP, and universal GRACE match the DFT data closely, instilling confidence that they will behave reliably at finite temperatures. The other potentials, by contrast, deviate from DFT markedly (Fig.~\ref{fig:phonons}).

\subsection{\label{subsec: P-V curves}Pressure-volume relation}

\begin{figure*}[!htb]
\begin{center}
\includegraphics[width=0.97999\textwidth]{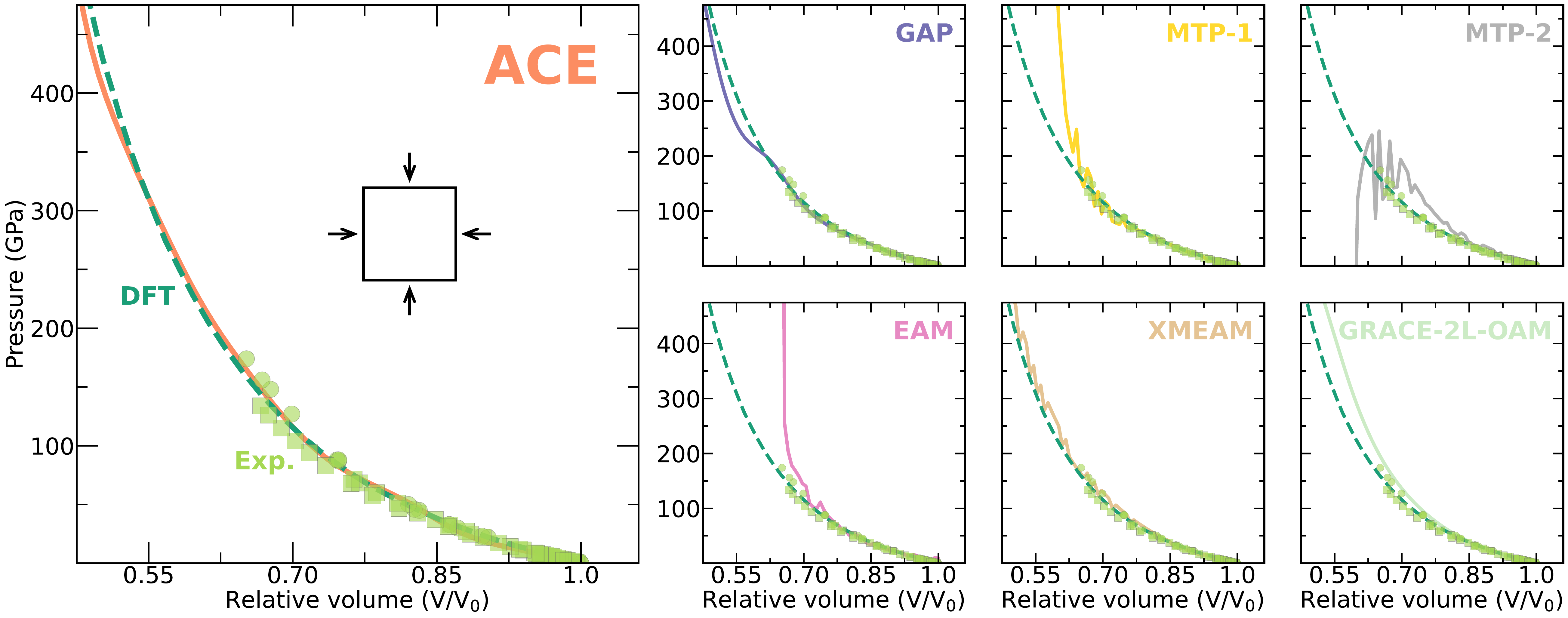}
\caption{Pressure-volume relation for the bcc Nb obtained with ACE and other interatomic potentials, compared with DFT and experimental data from Ref.~\cite{p_vs_v_dft_LANDA20062056} (dashed line), \cite{p_vs_v_exp_1_1970521} (circles), and~\cite{p_vs_v_exp_2_PhysRevB.73.224119} (squares). Ref.~\cite{p_vs_v_exp_2_PhysRevB.73.224119} also provides error bars.
}
\label{fig:p_vs_v}
\end{center}
\end{figure*} High pressures open the door to new physics~and~uncharted material properties~\cite{high_pressure_RevModPhys.90.015007}.~So~we~looked~at~how our ACE responds to pressure and obtained the pressure-volume relation, which interatomic potentials often struggle to reproduce~\cite{P_vs_V_Oleynik_10.1063/5.0218705}. ACE matches the experimental and DFT data across all considered pressures. This result, in tandem with accurate phonons (Fig.~\ref{fig:phonons}), evinces that our ACE could handle tasks such as the exploration of the niobium pressure-temperature phase diagram. In contrast, other potentials run afoul of DFT and experiment; ever larger they diverge as the pressure ascends. 

\subsection{\label{subsec: Vacancies}Vacancies}

Vacancies  govern self-diffusion in bcc metals~\cite{self_diff_1_PhysRevB.50.5928,self_diff_2_PhysRevB.80.144111}, shape the properties of irradiated materials~\cite{Gilbert_2008,Mason_2017,PRXEnergy.4.013008}, and much more~\cite{raguraman2025vacancyengineeringmetalsalloys}.
For diffusion, it is crucial to know both the vacancy formation energy and the barrier it must overcome to move to an adjacent site; the sum of the two constitutes the vacancy activation energy. We obtained all these quantities, including not only the migration barrier height but also its entire energy profile.

\begin{figure*}[!ht]
    \centering
    \includegraphics[width=0.97999\textwidth]{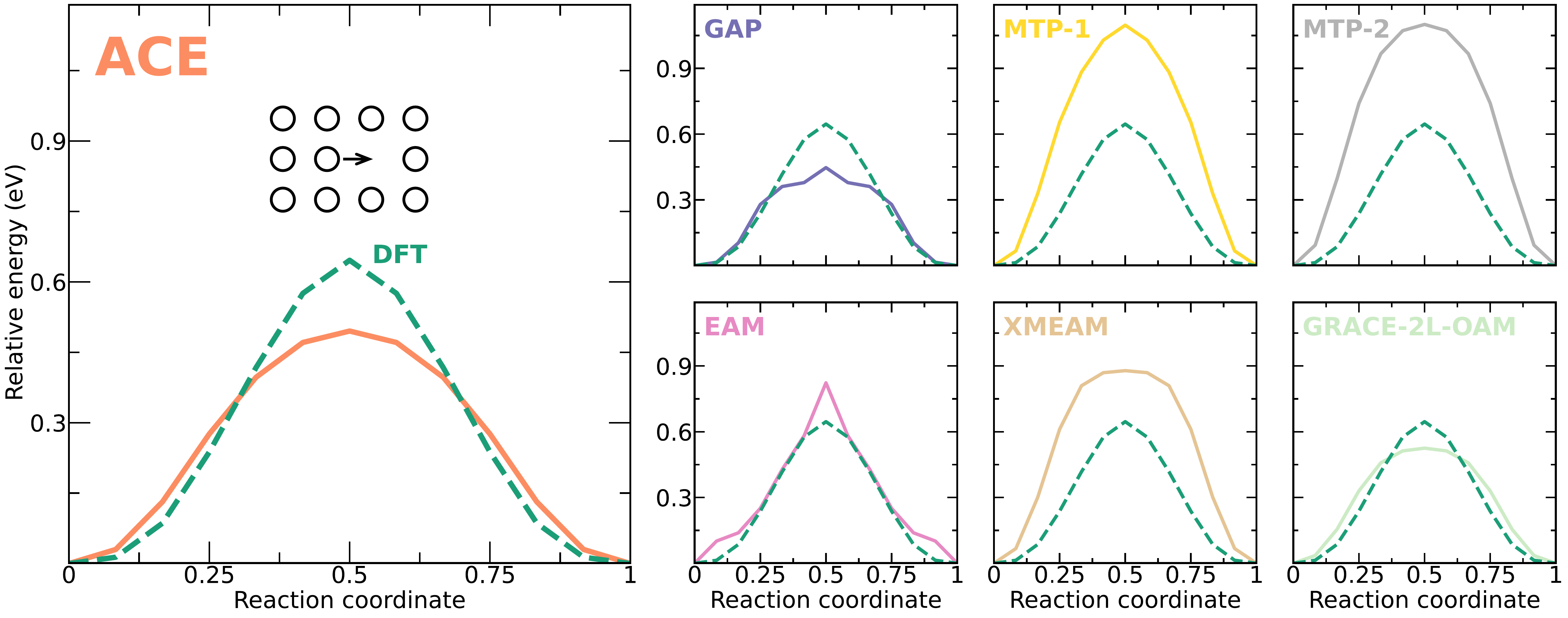}
    \caption{\label{fig:vacancies} Vacancy migration barriers in bcc Nb obtained with ACE and other interatomic potentials, compared with DFT~\cite{Dudarev_vacancy_PhysRevMaterials.3.063601}. }
  \end{figure*}

All ML-based potentials produce vacancy formation energies similar to the DFT values, while empirical potentials either underestimate (XMEAM) or overestimate it (EAM) (see Table~\ref{tab:vacancydata}). For migration barriers, only ACE and universal GRACE, although underestimating the height, are close to DFT (Fig.~\ref{fig:vacancies}). Accurate migration barriers led to accurate activation energies (Table~\ref{tab:vacancydata}), building confidence that, in finite-temperature simulations, atoms will diffuse at the proper rate.

\begin{table}[!htb]
\caption{\label{tab:vacancydata}
Vacancy formation energy ($E_V^{\mathrm{F}}$), vacancy migration energy ($E_V^{\mathrm{M}}$), and self-diffusion activation energy ($E^{\mathrm{SD}}$) obtained with ACE and other interatomic potentials, compared with DFT. $E^{\mathrm{SD}}$ is computed as $E_V^{\mathrm{F}} + E_V^{\mathrm{M}}$.}
\bgroup
\def\arraystretch{1.5}
\begin{ruledtabular}
\begin{tabular}{lccc}
 & $E_V^{\mathrm{F}}$ (eV) & $E_V^{\mathrm{M}}$ (eV) & $E^{\mathrm{SD}}$ (eV) \\
\hline
DFT         & 2.72\footnotemark[1], 2.77\footnotemark[2] & 0.65\footnotemark[1], 0.65\footnotemark[2] & 3.37\footnotemark[1], 3.42\footnotemark[2] \\
ACE         & 2.80       & 0.50       & 3.30       \\
GAP         & 2.85       & 0.45       & 3.30       \\
MTP-1       & 2.77       & 1.10       & 3.87       \\
MTP-2       & 2.78       & 1.10       & 3.88       \\
EAM         & 3.10       & 0.82       & 3.92       \\
XMEAM       & 2.43       & 0.88       & 3.31       \\
GRACE-2L-OAM & 2.74       & 0.53       & 3.27       \\
\end{tabular}
\end{ruledtabular}
\egroup
\footnotetext[1]{Ref. \cite{Dudarev_vacancy_PhysRevMaterials.3.063601}}
\footnotetext[2]{Ref. \cite{GAP_PhysRevLett.104.136403}}

\end{table}

\subsection{\label{subsec: E-V curves}Peierls barriers}

\begin{figure*}[!htb]
    \centering
    \includegraphics[width=0.97999\textwidth]{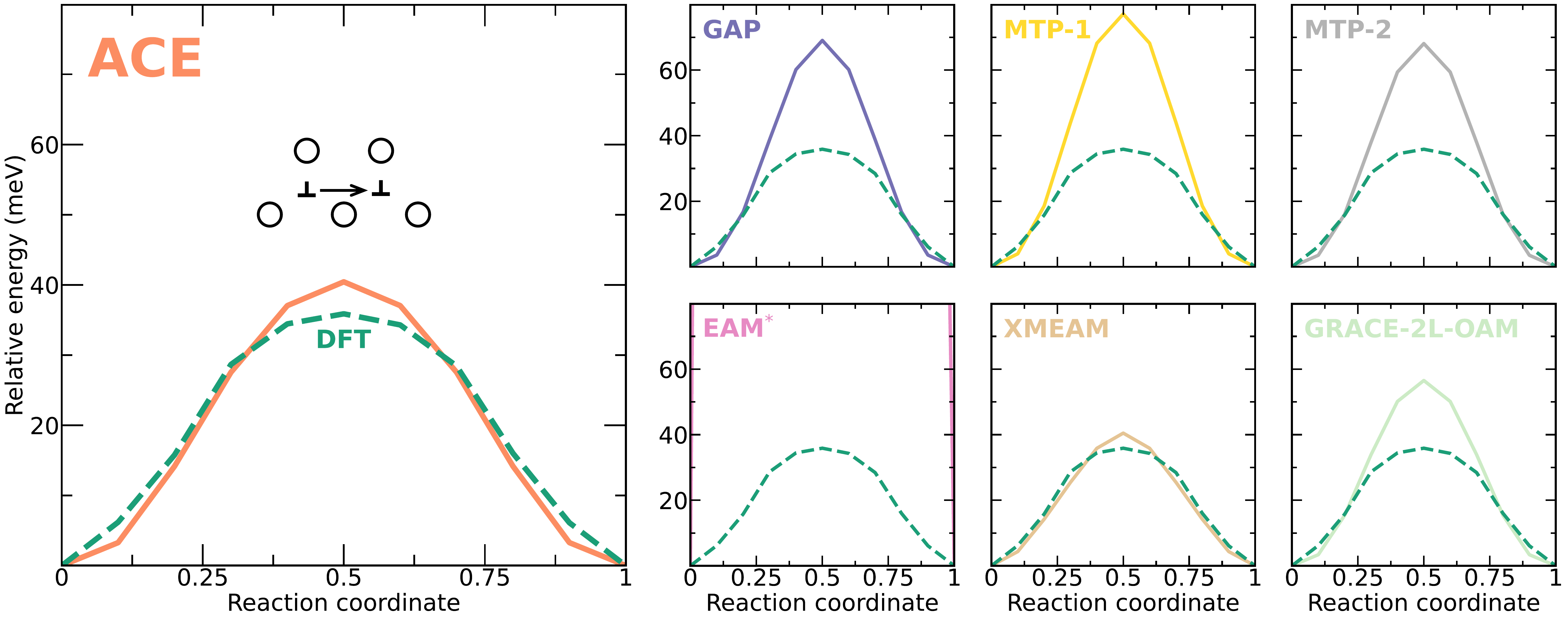}
    \caption{\label{fig:Peierls} Peierls barriers for $\frac{1}{2}$[111]~screw dislocation in bcc Nb obtained with ACE and other interatomic potentials, compared with DFT~\cite{Dezerald-dos-in-core}. ($*$) The calculation with the EAM potential did not converge; therefore, we report results after 1,000 relaxation steps (the peak---not visible in this graph---approaches 1000 meV). }
  \end{figure*}

Another defect, definitive for crystal properties, is dislocations~\cite{sutton2021concepts}. In bcc metals, $\frac{1}{2}$[111]~screw dislocations carry plastic \mbox{deformation~\cite{duesberyvitek1973corebcc, christian1983some, vitek2008non, weygand2015multiscale}}. We computed the energy profile of the barrier this dislocation must overcome during its glide—the Peierls barrier. \mbox{To get it, we} first determined the equilibrium dislocation configuration, characterized by the core structure~\cite{wang2022taming}. Aside from EAM, all potentials correctly produce the non-degenerate core (not shown). By contrast, only ACE and XMEAM produce the correct barriers (Fig.~\ref{fig:Peierls}); other potentials grossly overestimate the barrier height.  In simulations, this would lead to suppressed dislocation mobilities.

\subsection{\label{subsec: GBS}Grain boundary energies}
\begin{figure*}[!htb]
    \centering
    \includegraphics[width=0.97999\textwidth]{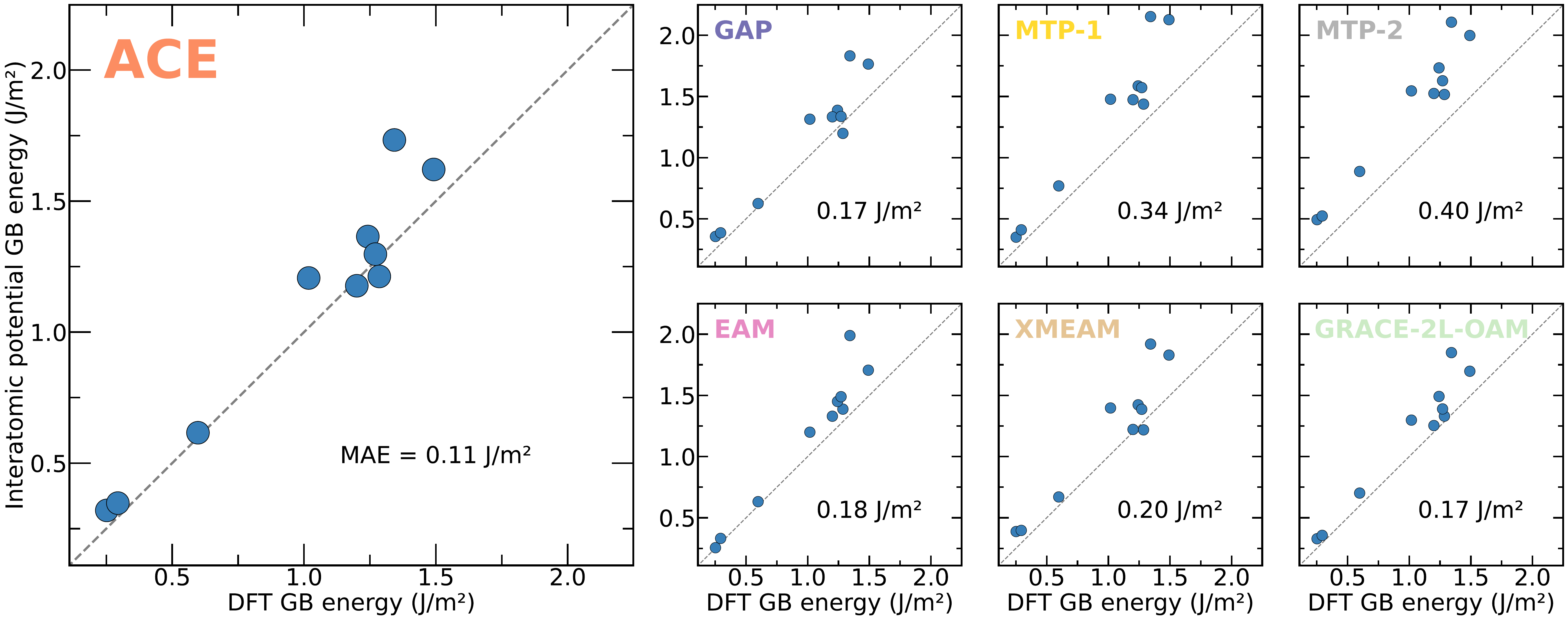}
    \caption{\label{fig:GBs}Parity plots for several grain boundary (GB) energies in bcc Nb obtained with ACE and six other interatomic potentials versus DFT data from the Grain Boundary Database (GBDB)~\cite{GB_energies_materials_project_ZHENG202040}, along with the corresponding mean absolute errors (MAE). 
    }
  \end{figure*}

The validation would not be complete without examining another crystal imperfection: the interfaces that seam grains of different orientations together—the grain boundaries (GBs). GBs hold sway over a wide variety of processes, including plasticity, fracture, and diffusion~\cite{GBs_review_Dehm2022,GBs_diffus_Balluffi1982}. We looked at how well our ACE and other potentials capture the energies of several coincidence-site-lattice GBs. 
Apart from two MTPs, all potentials produce accurate GB energies, with ACE—based on mean absolute error (MAE)—coming out on top (Fig.~\ref{fig:GBs}).

\subsection{\label{subsec:comp_speed}Computational speed}

\begin{figure}[!htb]
\begin{center}
\includegraphics[width=0.401\textwidth]{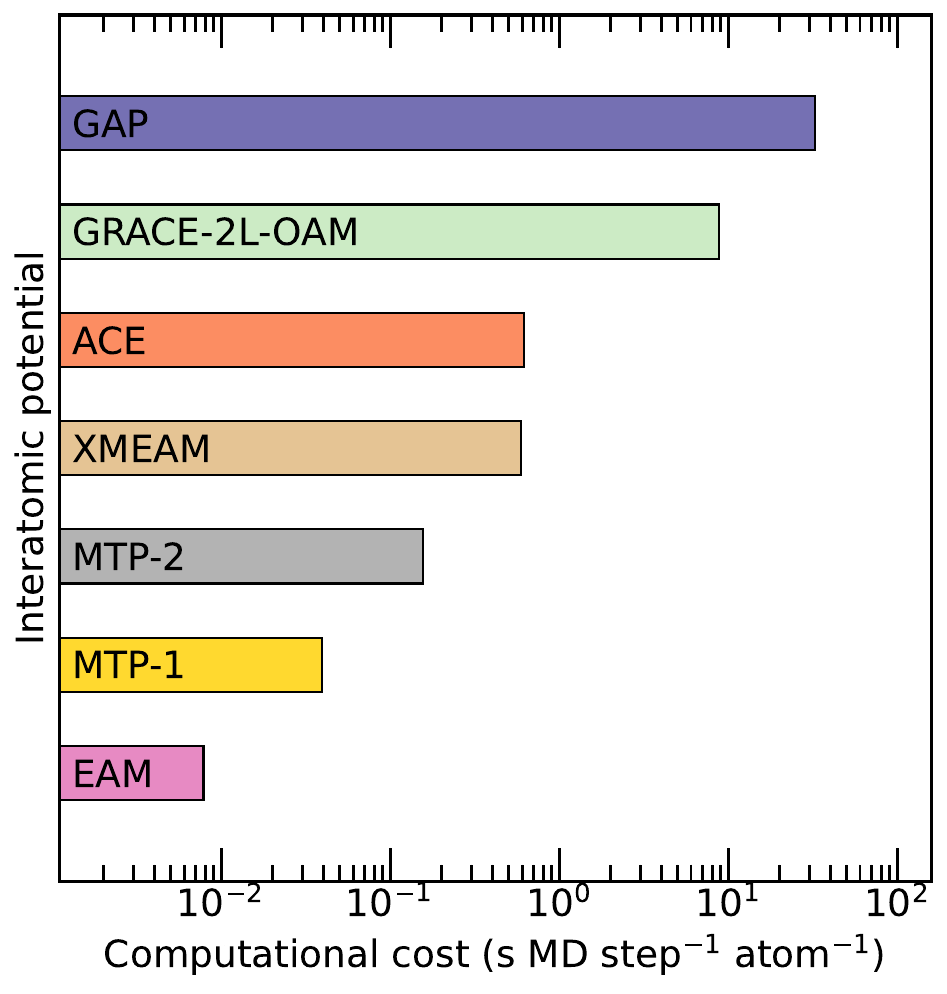}
\caption{Computational speed of ACE and other potentials.
}
\label{fig:comp_speed}
\end{center}
\end{figure}

Along with accuracy, computational speed is a key ingredient of a useful interatomic potential. Therefore, we gauged the speed of the potentials in molecular dynamics simulations (see Appendix~\ref{ap:valid_details} for details). Although our ACE is not the fastest, it is still fast, being two orders of magnitude more efficient than GAP and on par with the empirical XMEAM (Fig.~\ref{fig:comp_speed}). The potentials boosting the highest speed—EAM and two MTPs—do not, as we laid bare above, furnish the desired accuracy. Therefore, in simulations, this speed may turn out to be futile.

\section{\label{sec:appl}Application to fracture}

In the following, we showcase that our new ACE delivers not only near-DFT accuracy but also the capacity for large-scale simulations. The exemplary application should (1) demand simulation cells with hundreds of thousands or even millions of atoms and (2) simultaneously be shaped by atomic interactions. The epitome of such an application is fracture.

Fracture implies crack growth, unfolding where stresses concentrate---at the crack tip. \mbox{In brittle crystals, under} load, bonds between atoms at the very crack tip break; the crack tip cleaves and grows~\cite{Lawn1980_sharp_Si_crack}. Despite the local nature of these events, the simulation \mbox{cell must be} large enough to prevent the long-range stress field around the crack tip from interacting with the cell boundary~\cite{andric_curtin_2018atomistic}.

\begin{figure*}[!htb]
    \centering
    \includegraphics[width=0.99999\textwidth]{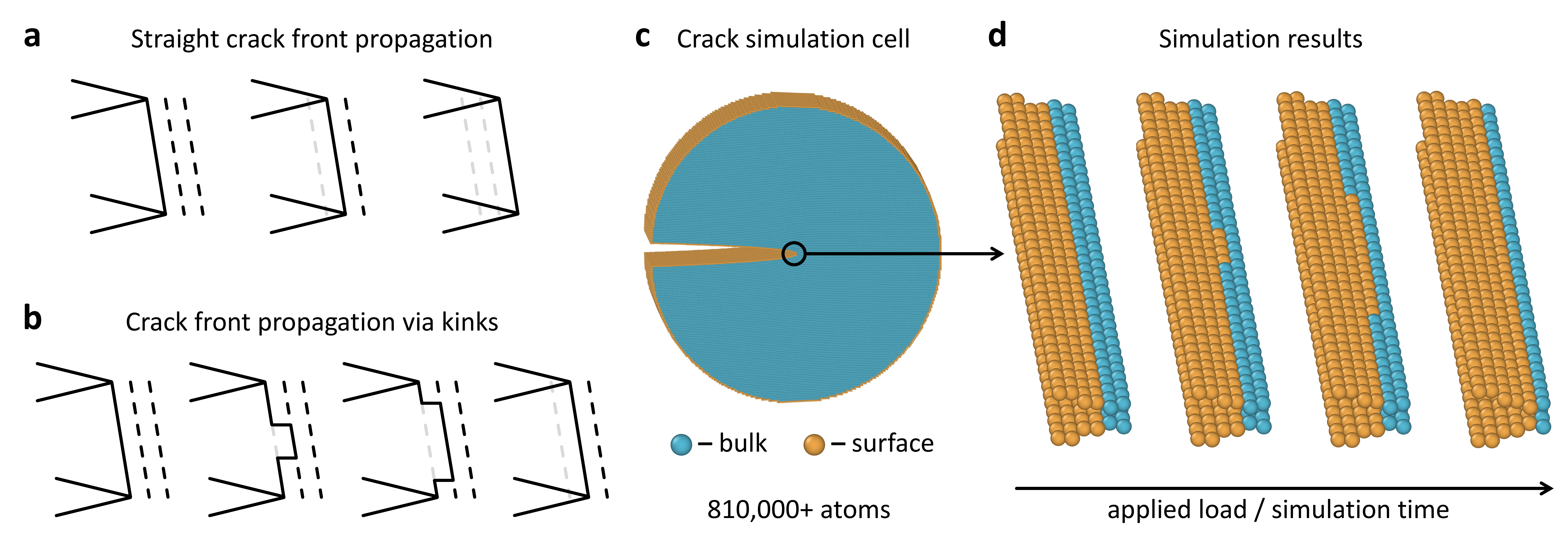}
    \caption{\label{fig:crack} (a-b) Schematic of crack-front propagation as a straight, planar front (a) and a front advancing by kink nucleation and growth (b). (c) Simulation cell employed to scrutinize crack propagation; the cell's radius is 200~Å and width is 117~Å. (d)~4.2~K crack-propagation simulation revealing the crack front advancing via kink nucleation and growth.}
  \end{figure*}

The cell must also be thick enough. One might picture the crack front propagating as a straight line with bonds across the front breaking all at once (Fig.~\ref{fig:crack}a). But that is not what is happening. It is much more energetically favorable when bonds break locally, so that only a small section of the crack front, called a double kink, propagates at first~\cite{kinks_1971_10.1063/1.1660757,thomson_kinks_1973_10.1063/1.1662512,kermode_kinks_PhysRevLett.115.135501,kinks_exp_Cochard2024,kinks_explained_Marder2024}; the double kink then grows laterally until the whole crack front end up at the next position (Fig.~\ref{fig:crack}b); the process then repeats. Therefore, the simulation cell must be thick enough to accommodate kink nucleation and growth.

To mimic the above crack propagation, \mbox{we employed a} \mbox{117-Å-thick} cylindrical simulation cell with a \mbox{200-Å}~radius, containing over 810,000 atoms (Fig.~\ref{fig:crack}c). We then simulated niobium fracture at liquid helium temperature of 4.2~K, where, based on experimental observations~\cite{nb_4.2_K_KubinJouffrey1971,nb_4.2_K_Walsh2002NbFracture}, brittle crack propagation is expected.

As the simulation began, the crack front started to tremble. After a little, a small kink pair appeared at the center of the front; soon it grew to the left and to the right, met its periodic images, and the full crack front propagated---a physically sound behavior that our ACE has captured (Fig.~\ref{fig:crack}d).

\section{\label{sec:discussion}Summary and conclusions}

We developed a general-purpose atomic cluster expansion (ACE) interatomic potential for niobium. Expanding the existing DFT training database~\cite{GAP_Nb_PhysRevMaterials.4.093802} ensured ACE stability. Validation across a wide range of properties—phonons, the pressure-volume relation, the Peierls barrier for screw dislocations, and many others—along with comparisons to existing potentials, showed that only our ACE captures the DFT ground truth across all tests. Despite the accuracy of the universal GRACE potential~\cite{lysogorskiy2025graphatomicclusterexpansion}, ACE still performed better overall (and is much cheaper to run), underscoring the utility of bespoke potentials (see also Ref.~\cite{Pickard_bespoke_utility} or Appendix A13 of Ref.~\cite{foundation_model_10.1063/5.0297006}). In an \mbox{810,000-atom} fracture simulation, we demonstrated ACE's capacity for large-scale applications. Balancing accuracy, speed, and robustness, our ACE is paving the way for exploring niobium in large-scale atomistic simulations with \mbox{near-DFT} precision.  

\section*{Acknowledgment}

We thank Mike Finnis for a valuable discussion of phonon anomalies in bcc niobium and Yury Lysogorskiy for technical support. We acknowledge Rui Wang for sharing the XMEAM potential and for providing instructions for its use in simulations. We thank Matous Mrovec for his valuable insights into the results.

\section*{Data availability}

The Nb ACE potential, its training input file, and the expansions to the DFT reference database are publicly available~\cite{data-avail-comment}. \mbox{Step-by-step} instructions on how to use ACE potentials in \texttt{LAMMPS}~\cite{LAMMPS} or as an \texttt{ASE} calculator~\cite{larsen2017ase} are available on the {\UrlFont{pacemaker}} website~\cite{pmusefitpot}

\appendix

\section{\label{ap:comput_det}Computational details}

\subsection{Density Functional Theory (DFT) calculations}

DFT calculations were performed with settings identical to those in the original GAP work~\cite{GAP_Nb_PhysRevMaterials.4.093802}, which~provided the data to train our ACE. We used the {\UrlFont{VASP}} package~\cite{vasp-1,vasp-2,vasp-3} with the projector~augmented-wave~(PAW) method~\cite{PAW}, the PBE generalized gradient approximation (GGA-PBE) for exchange–correlation~\cite{PBE}, a plane-wave energy cutoff of 500 eV, and a $k$-point spacing of 0.15~\AA$^{-1}$. Spin-polarization was included, and Gaussian smearing with a width of 0.1 eV was applied.

\subsection{Atomic cluster expansion (ACE) training}

We trained our ACE using the {\UrlFont{pacemaker}} package~\cite{pacemaker-Lysogorskiy2021}. Our nonlinear ACE employs the Finnis-Sinclair embedding. We selected a cutoff of 8~Å, determined through trial and error, to balance precision and speed. In training, we assigned relative weights of 0.7 to energies and 0.3 to forces. This relatively high weight for the forces was justified by the accurate phonon spectra obtained from the resulting model. A complete set of hyperparameters can be found in the {\UrlFont{pacemaker}} input file~\cite{data-avail-comment}.

\subsection{\label{ap:valid_details}Validation details and setups}

For validation, we created a Python workflow based on the Atomic Simulation Environment ({\UrlFont{ASE}})~\cite{larsen2017ase}. We computed the elastic constants, vacancy formation energies, and phonons with the {\UrlFont{amstools}} package~\cite{amstools}.
The {\UrlFont{phonopy}} package~\cite{phonopy}, integrated within {\UrlFont{amstools}}, was used for phonon calculations. We used the nudged elastic band (NEB) method~\cite{NEB_10.1063/1.1329672}, \mbox{as implemented in {\UrlFont{ASE}}, to} evaluate Peierls barriers and vacancy migration barriers. Screw dislocations for Peierls barriers were created using the {\UrlFont{babel}} package~\cite{babel}.
We applied a 0.01~eV/Å criterion for atomic relaxations, except for grain boundaries, where it was 0.02~eV/Å---both to align with the DFT data we compared against. \mbox{To get the E-V} curves, the energies were normalized to ensure all potentials match the minimum energy of the most stable bcc phase.

To assess computational speed in molecular dynamics (MD) simulations, we used the {\UrlFont{LAMMPS}} package~\cite{LAMMPS} to perform 1000 MD steps in the NVT ensemble at 100~K. The simulations were performed on an Intel~Xeon~E5-1620~v4~CPU~@~3.5~GHz, using a 4×4×4 bcc simulation cell with 128 atoms. The MD time step was set to 3 fs.

We used eight-layer slabs to compute surface energies and traction-separation (T-S) curves. For phonons, a 6×6×6 supercell with 412 atoms was employed, and for vacancy formation energies and migration barriers, a 3×3×3 supercell with 54 atoms. These supercell sizes were chosen to maintain consistency with the DFT calculations against which we compared our results.

To compute the Peierls barriers, we introduced two $\frac{1}{2}$[111]~screw dislocations with opposite Burgers vectors into a simulation cell, so they form a quadrupolar arrangement~\cite{ventelon2007corequadr, PhysRevLett.102.055502-Clouet-2009, Clouet2020-ab-initio}. The dislocations were introduced using anisotropic elasticity theory as implemented in the~{\UrlFont{babel}} package~\cite{babel}. Homogeneous strain was applied to the simulation cell to offset the strain caused by the dislocations~\cite{Clouet2020-ab-initio, Comptes_Rendus_Physique_2021_clouet, Kraych2019npj}. The simulation cell contained 135 atoms, matching the size used in the DFT study served as our benchmark. To create a second, final dislocation position for the NEB calculations, we replicated the above procedure for the initial position, shifting both dislocations within the simulation cell to adjacent Peierls valleys. This process was conducted strictly following the procedure outlined in the DFT benchmark study~\cite{Dezerald-dos-in-core}.

DFT grain boundary (GB) supercells were taken from the Grain Boundary Database (GBDB)~\cite{GB_energies_materials_project_ZHENG202040}, which is part of the Materials Project~\cite{Jain2013materialsproject}. 
These supercells were scaled to match the lattice parameters obtained with different potentials, and the atomic positions were then relaxed. GB energies, $\gamma_\mathrm{GB}$, were determined as
\begin{align*}\label{eq:EGB}
\gamma_\mathrm{GB} = \frac{E_\mathrm{GB} - E_\mathrm{bulk}}{2A_\mathrm{GB}},
\end{align*}
\noindent where $E_\mathrm{GB}$ and $E_\mathrm{bulk}$ are the total energies of a supercell with the GB and of a bulk supercell with the same number of atoms. $A_\mathrm{GB}$ is the GB area; \mbox{the 2 in} the denominator accounts for two GBs in the periodic supercell.

\subsection{\label{ap:appl_details}Fracture simulations details and setups}

To model fracture, we employed a cylindrical simulation cell containing a pre-existing crack on one side, with the crack tip located at the center (Fig.~\ref{fig:crack}c). Loading was applied via the mode-I stress intensity factor, $K_{\mathrm{I}}$, which uniquely determines the stress and strain fields around the crack tip~\cite{RITCHIE2021_chap_3}. For each value of $K_{\mathrm{I}}$, all atoms were displaced according to the corresponding linear elastic strain field, while atoms within a \mbox{5~\AA-thick} outer boundary were held fixed. When the cell radius is converged this simulation setup obeys linear elastic fracture mechanics (LEFM)~\cite{andric_curtin_2018atomistic}, and LEFM, in turn, has been experimentally validated~\cite{kitamura_lefm_exp_1_sumigawa2017griffith,kitamura_lefm_exp_2_gallo2018fracture}. Ref.~\cite{egorov2025crackcrackhydrogenfavors} provides more details on implementing this approach.

We explored the (110)[110] crack system, with (110) the crack plane and [110] the crack front direction. The cell radius was set to 200~\AA{} based on a convergence test of the critical $K_{\mathrm{I}}$ for cleavage. In this test, we set up a 2D system with a thickness of 4.7~\AA. For the simulation displayed in Fig.~\ref{fig:crack}, we set up a 3D system with a cell thickness of 117~\AA{} to accommodate kink-pair formation. We performed molecular dynamics (MD) fracture simulations with the {\UrlFont{LAMMPS}} package in the NVT ensemble using a time step of 3.5 fs. This time step is roughly 1/45 of the period of the highest phonon frequency~\cite{MD_timestep_KIM201460}. A loading rate of $2.9 \times 10^{8}\,\mathrm{MPa}\sqrt{\mathrm{m}}\,\mathrm{s}^{-1}$ was used. The starting $K_{\mathrm{I}}$ was $0.630~\mathrm{MPa}\sqrt{\mathrm{m}}$, and the crack front fully propagated just before $K_{\mathrm{I}}$ reached $0.645~\mathrm{MPa}\sqrt{\mathrm{m}}$. The crack front was visualized using common-neighbour analysis~\cite{CNA-1} implemented in the {\UrlFont{OVITO}} package~\cite{OVITOstukowski2009visualization}.

\bibliography{bibliography}

@article{nb_4.2_K_KubinJouffrey1971,
  author  = {Kubin, L. P. and Jouffrey, B.},
  title   = {On low temperature plastic instability in pure niobium single crystals},
  journal = {Philos. Mag.},
  volume  = {24},
  number  = {188},
  pages   = {437--449},
  year    = {1971},
  doi     = {10.1080/14786437108227399},
  url     = {https://doi.org/10.1080/14786437108227399}
}

@article{Beyerlein_mlpot_WANG2021110364,
title = {Generalized stacking fault energies and Peierls stresses in refractory body-centered cubic metals from machine learning-based interatomic potentials},
journal = {Comput. Mater. Sci.},
volume = {192},
pages = {110364},
year = {2021},
issn = {0927-0256},
doi = {https://doi.org/10.1016/j.commatsci.2021.110364},
url = {https://www.sciencedirect.com/science/article/pii/S0927025621000896},
author = {Xiaowang Wang and Shuozhi Xu and Wu-Rong Jian and Xiang-Guo Li and Yanqing Su and Irene J. Beyerlein},
}

@inproceedings{nb_4.2_K_Walsh2002NbFracture,
  author    = {Walsh, R. P. and Han, Ke and Toplosky, V. J. and Mitchell, R. R.},
  title     = {Low temperature tensile and fracture characteristics of high purity niobium},
  booktitle = {AIP Conf. Proc.},
  volume    = {614},
  number    = {1},
  pages     = {186--196},
  year      = {2002},
  doi       = {10.1063/1.1472542},
  url       = {https://doi.org/10.1063/1.1472542}
}

@article{kinks_explained_Marder2024,
  author  = {Marder, Michael},
  title   = {Breaking fast and slow},
  journal = {Nat. Phys.},
  volume  = {20},
  number  = {4},
  pages   = {546--547},
  year    = {2024},
  doi     = {10.1038/s41567-024-02389-0},
  url     = {https://doi.org/10.1038/s41567-024-02389-0}
}

@article{kinks_exp_Cochard2024,
  author  = {Cochard, T.  and others},
  title   = {Propagation of extended fractures by local nucleation and rapid transverse expansion of crack-front distortion},
  journal = {Nat. Phys.},
  volume  = {20},
  number  = {4},
  pages   = {660--665},
  year    = {2024},
  doi     = {10.1038/s41567-023-02365-0},
  url     = {https://doi.org/10.1038/s41567-023-02365-0}
}

@article{kinks_1971_10.1063/1.1660757,
  author  = {Gilman, J. J. and Tong, H. C.},
  title   = {Quantum Tunneling as an Elementary Fracture Process},
  journal = {J. Appl. Phys.},
  volume  = {42},
  number  = {9},
  pages   = {3479--3486},
  year    = {1971},
  doi     = {10.1063/1.1660757},
  url     = {https://doi.org/10.1063/1.1660757}
}

@article{thomson_kinks_1973_10.1063/1.1662512,
  author  = {Hsieh, C. and Thomson, R.},
  title   = {Lattice theory of fracture and crack creep},
  journal = {J. Appl. Phys.},
  volume  = {44},
  number  = {5},
  pages   = {2051--2063},
  year    = {1973},
  doi     = {10.1063/1.1662512},
  url     = {https://doi.org/10.1063/1.1662512}
}

@misc{pmusefitpot,
  title        = {Use Fitted Potential},
  year         = {2026},
  howpublished = {\url{https://pacemaker.readthedocs.io/en/latest/pacemaker/quickstart/\#using_fitted_potential}},
  note         = {Accessed: 2026-05-14}
}

@misc{egorov2025crackcrackhydrogenfavors,
      title={To crack, or not to crack: How hydrogen favors crack propagation in iron at the atomic scale}, 
      author={Aleksei Egorov and Lei Zhang and Erik van der Giessen and Francesco Maresca},
      year={2025},
      eprint={2512.12843},
      archivePrefix={arXiv},
      primaryClass={cond-mat.mtrl-sci},
      doi={https://doi.org/10.48550/arXiv.2512.12843},
      url={https://arxiv.org/abs/2512.12843}, 
}

@article{kermode_kinks_PhysRevLett.115.135501,
  title = {Low Speed Crack Propagation via Kink Formation and Advance on the Silicon (110) Cleavage Plane},
  author = {Kermode, James R. and Gleizer, Anna and Kovel, Guy and Pastewka, Lars and Cs\'anyi, G\'abor and Sherman, Dov and De Vita, Alessandro},
  journal = {Phys. Rev. Lett.},
  volume = {115},
  issue = {13},
  pages = {135501},
  numpages = {5},
  year = {2015},
  month = {Sep},
  publisher = {American Physical Society},
  doi = {10.1103/PhysRevLett.115.135501},
  url = {https://link.aps.org/doi/10.1103/PhysRevLett.115.135501}
}

@article{MISHIN2021116980,
title = {Machine-learning interatomic potentials for materials science},
journal = {Acta Mater.},
volume = {214},
pages = {116980},
year = {2021},
issn = {1359-6454},
doi = {https://doi.org/10.1016/j.actamat.2021.116980},
url = {https://www.sciencedirect.com/science/article/pii/S1359645421003608},
author = {Y. Mishin},
}

@article{roadmap_Zhang_2025,
doi = {10.1088/1361-651X/ad9d63},
url = {https://doi.org/10.1088/1361-651X/ad9d63},
year = {2025},
month = {jan},
publisher = {IOP Publishing},
volume = {33},
number = {2},
pages = {023301},
author = {Zhang, Yong-Wei and others},
title = {Roadmap for the development of machine learning-based interatomic potentials},
journal = {Model. Simul. Mater. Sci. Eng.}
}

@article{Jain2013materialsproject,
doi = {10.1063/1.4812323},
issn = {2166532X},
journal = {APL Mater.},
number = {1},
pages = {011002},
title = {{The Materials Project: A materials genome approach to accelerating materials innovation}},
url = {http://link.aip.org/link/AMPADS/v1/i1/p011002/s1\&Agg=doi},
volume = {1},
year = {2013},
author = {Jain, Anubhav and Ong, Shyue Ping and Hautier, Geoffroy and Chen, Wei and Richards, William Davidson and Dacek, Stephen and Cholia, Shreyas and Gunter, Dan and Skinner, David and Ceder, Gerbrand and Persson, Kristin Aslaug}
}

@article{GB_energies_materials_project_ZHENG202040,
title = {Grain boundary properties of elemental metals},
journal = {Acta Mater.},
volume = {186},
pages = {40-49},
year = {2020},
issn = {1359-6454},
doi = {https://doi.org/10.1016/j.actamat.2019.12.030},
url = {https://www.sciencedirect.com/science/article/pii/S1359645419308699},
author = {Hui Zheng and Xiang-Guo Li and Richard Tran and Chi Chen and Matthew Horton and Donald Winston and Kristin Aslaug Persson and Shyue Ping Ong}
}

@article{xmeam_vanadium_PhysRevMaterials.6.113603,
  title = {Classical and machine learning interatomic potentials for BCC vanadium},
  author = {Wang, Rui and Ma, Xiaoxiao and Zhang, Linfeng and Wang, Han and Srolovitz, David J. and Wen, Tongqi and Wu, Zhaoxuan},
  journal = {Phys. Rev. Mater.},
  volume = {6},
  issue = {11},
  pages = {113603},
  numpages = {23},
  year = {2022},
  month = {Nov},
  publisher = {American Physical Society},
  doi = {10.1103/PhysRevMaterials.6.113603},
  url = {https://link.aps.org/doi/10.1103/PhysRevMaterials.6.113603}
}

@Article{GBs_diffus_Balluffi1982,
author={Balluffi, R. W.},
title={Grain boundary diffusion mechanisms in metals},
journal={Metall. Trans. B},
year={1982},
month={Dec},
day={01},
volume={13},
number={4},
pages={527-553},
issn={2379-0229},
doi={10.1007/BF02650011},
url={https://doi.org/10.1007/BF02650011}
}

@Article{GBs_review_Dehm2022,
author={Dehm, Gerhard
and Cairney, Julie},
title={Implication of grain-boundary structure and chemistry on plasticity and failure},
journal={MRS Bull.},
year={2022},
month={Aug},
day={01},
volume={47},
number={8},
pages={800-807},
issn={1938-1425},
doi={10.1557/s43577-022-00378-3},
url={https://doi.org/10.1557/s43577-022-00378-3}
}

@article{Leimeroth_Albe_comparison_2025,
doi = {10.1088/1361-651X/adf56d},
url = {https://doi.org/10.1088/1361-651X/adf56d},
year = {2025},
month = {aug},
publisher = {IOP Publishing},
volume = {33},
number = {6},
pages = {065012},
author = {Leimeroth, Niklas and Erhard, Linus C and Albe, Karsten and Rohrer, Jochen},
title = {Machine-learning interatomic potentials from a users perspective: a comparison of accuracy, speed and data efficiency},
journal = {Model. Simul. Mater. Sci. Eng.}
}

@article{emp_pots_limit_Beyerlein,
author = {R.F. Zhang and J. Wang and I.J. Beyerlein and T.C. Germann},
title = {Twinning in bcc metals under shock loading: a challenge to empirical potentials},
journal = {Philos. Mag. Lett.},
volume = {91},
number = {12},
pages = {731--740},
year = {2011},
publisher = {Taylor \& Francis},
doi = {10.1080/09500839.2011.615348},
URL = {
https://doi.org/10.1080/09500839.2011.615348
},
}

@article{Pickard_bespoke_utility,
author = {Conway, Lewis J. and Pickard, Chris J.},
title = {Accelerating Crystal Structure Prediction Using Data-Derived Potentials: High-Pressure Binary Hydrides},
journal = {Ann. Phys.},
volume = {538},
number = {3},
pages = {e00608},
doi = {https://doi.org/10.1002/andp.202500608},
url = {https://onlinelibrary.wiley.com/doi/abs/10.1002/andp.202500608},
year = {2026}
}

@article{foundation_model_10.1063/5.0297006,
    author = {Batatia, Ilyes and others},
    title = {A foundation model for atomistic materials chemistry},
    journal = {J. Chem. Phys.},
    volume = {163},
    number = {18},
    pages = {184110},
    year = {2025},
    month = {11},
    issn = {0021-9606},
    doi = {10.1063/5.0297006},
    url = {https://doi.org/10.1063/5.0297006},
}

@article{emp_pot_limit_RODNEY2008418,
title = {Atomic modeling of irradiation-induced hardening},
journal = {C. R. Phys.},
volume = {9},
number = {3},
pages = {418-426},
year = {2008},
note = {{Materials subjected to fast neutron irradiation}},
issn = {1631-0705},
doi = {https://doi.org/10.1016/j.crhy.2007.08.005},
url = {https://www.sciencedirect.com/science/article/pii/S1631070507002253},
author = {David Rodney},
}

@misc{raguraman2025vacancyengineeringmetalsalloys,
      title={{Vacancy Engineering in Metals and Alloys}}, 
      author={Sreenivas Raguraman and Homero Reyes Pulido and Christopher Hutchinson and Arun Devaraj and Marc H. Weber and Michael L. Falk and Timothy P. Weihs},
      year={2025},
      eprint={2511.20706},
      archivePrefix={arXiv},
      primaryClass={cond-mat.mtrl-sci},
      url={https://arxiv.org/abs/2511.20706}, 
}

@article{nb_discovery_10.1098/rstl.1802.0005,
    author = {Hatchett, Charles},
    title = {III. An analysis of a mineral substance from North America, containing a metal bitberto unknown},
    journal = {Philos. Trans. R. Soc. London},
    number = {92},
    pages = {49-66},
    year = {1802},
    month = {12},
    issn = {0261-0523},
    doi = {10.1098/rstl.1802.0005},
    url = {https://doi.org/10.1098/rstl.1802.0005}
}

@misc{gap_data_nb_training,
  author       = {{Byggmästar~$et$~$al.$}},
  title        = {{GAP DFT reference database for niobium}},
  year         = {2019},
  howpublished = {\url{https://gitlab.com/acclab/gap-data/-/tree/master/Nb/training-data}},
  publisher    = {GitLab}
}

@Article{lysogorskiy2025graphatomicclusterexpansion,
author={Lysogorskiy, Yury
and Bochkarev, Anton
and Drautz, Ralf},
title={Graph atomic cluster expansion for foundational machine learning interatomic potentials},
journal={npj Comput. Mater.},
year={2026},
month={Feb},
day={08},
url={https://doi.org/10.1038/s41524-026-01979-1}
}

@article{Grabowski_origin_XU2024120423,
title = {{Origin of the yield stress anomaly in L12 intermetallics unveiled with physically informed machine-learning potentials}},
journal = {Acta Mater.},
volume = {281},
pages = {120423},
year = {2024},
issn = {1359-6454},
doi = {https://doi.org/10.1016/j.actamat.2024.120423},
url = {https://www.sciencedirect.com/science/article/pii/S1359645424007730},
author = {Xiang Xu and Xi Zhang and Erik Bitzek and Siegfried Schmauder and Blazej Grabowski}
}

@incollection{p_vs_v_exp_1_1970521,
title = {APPENDIX D - SHOCK WAVE DATA FOR SOLIDS},
editor = {RAY KINSLOW},
booktitle = {High-Velocity Impact Phenomena},
publisher = {Academic Press},
pages = {521-529},
year = {1970},
isbn = {978-0-12-408950-1},
doi = {https://doi.org/10.1016/B978-0-12-408950-1.50018-5},
url = {https://www.sciencedirect.com/science/article/pii/B9780124089501500185}
}

@article{p_vs_v_exp_2_PhysRevB.73.224119,
  title = {{High-pressure equation of state for Nb with a helium-pressure medium: Powder x-ray diffraction experiments}},
  author = {Kenichi, Takemura and Singh, Anil K.},
  journal = {Phys. Rev. B},
  volume = {73},
  issue = {22},
  pages = {224119},
  numpages = {9},
  year = {2006},
  month = {Jun},
  publisher = {American Physical Society},
  doi = {10.1103/PhysRevB.73.224119},
  url = {https://link.aps.org/doi/10.1103/PhysRevB.73.224119}
}

@article{p_vs_v_dft_LANDA20062056,
title = {{Ab initio calculations of elastic constants of the bcc V–Nb system at high pressures}},
journal = {J. Phys. Chem. Solids},
volume = {67},
number = {9},
pages = {2056-2064},
year = {2006},
issn = {0022-3697},
doi = {https://doi.org/10.1016/j.jpcs.2006.05.027},
url = {https://www.sciencedirect.com/science/article/pii/S0022369706002964},
author = {A. Landa and J. Klepeis and P. Söderlind and I. Naumov and O. Velikokhatnyi and L. Vitos and A. Ruban}
}

@article{high_pressure_RevModPhys.90.015007,
  title = {Solids, liquids, and gases under high pressure},
  author = {Mao, Ho-Kwang and Chen, Xiao-Jia and Ding, Yang and Li, Bing and Wang, Lin},
  journal = {Rev. Mod. Phys.},
  volume = {90},
  issue = {1},
  pages = {015007},
  numpages = {55},
  year = {2018},
  month = {Mar},
  publisher = {American Physical Society},
  doi = {10.1103/RevModPhys.90.015007},
  url = {https://link.aps.org/doi/10.1103/RevModPhys.90.015007}
}

@article{P_vs_V_Oleynik_10.1063/5.0218705,
    author = {Willman, Jonathan T. and Gonzalez, Joseph M. and Nguyen-Cong, Kien and Hamel, Sebastien and Lordi, Vincenzo and Oleynik, Ivan I.},
    title = {Accuracy, transferability, and computational efficiency of interatomic potentials for simulations of carbon under extreme conditions},
    journal = {J. Chem. Phys.},
    volume = {161},
    number = {8},
    pages = {084709},
    year = {2024},
    month = {08},
    issn = {0021-9606},
    doi = {10.1063/5.0218705},
    url = {https://doi.org/10.1063/5.0218705}
}

@article{PRXEnergy.4.013008,
  title = {Unraveling Temperature-Induced Vacancy Clustering in Tungsten: From Direct Microscopy to Atomistic Insights via Data-Driven Bayesian Sampling},
  author = {Zhong, Anruo and Lapointe, Clovis and Goryaeva, Alexandra M. and Arakawa, Kazuto and Ath\`enes, Manuel and Marinica, Mihai-Cosmin},
  journal = {PRX Energy},
  volume = {4},
  issue = {1},
  pages = {013008},
  numpages = {16},
  year = {2025},
  month = {Feb},
  publisher = {American Physical Society},
  doi = {10.1103/PRXEnergy.4.013008},
  url = {https://link.aps.org/doi/10.1103/PRXEnergy.4.013008}
}

@Article{Kormann_Kostiuchenko2019,
author={Kostiuchenko, Tatiana
and K{\"o}rmann, Fritz
and Neugebauer, J{\"o}rg
and Shapeev, Alexander},
title={Impact of lattice relaxations on phase transitions in a high-entropy alloy studied by machine-learning potentials},
journal={npj Comput. Mater.},
year={2019},
month={May},
day={01},
volume={5},
number={1},
pages={55},
issn={2057-3960},
doi={10.1038/s41524-019-0195-y},
url={https://doi.org/10.1038/s41524-019-0195-y}
}

@Article{Kraych2019npj,
author={Kraych, Antoine
and Clouet, Emmanuel
and Dezerald, Lucile
and Ventelon, Lisa
and Willaime, Fran{\c{c}}ois
and Rodney, David},
title={Non-glide effects and dislocation core fields in BCC metals},
journal={npj Comput. Mater.},
year={2019},
month={Nov},
day={14},
volume={5},
number={1},
pages={109},
issn={2057-3960},
doi={10.1038/s41524-019-0247-3},
url={https://doi.org/10.1038/s41524-019-0247-3}
}

@article{Comptes_Rendus_Physique_2021_clouet,
     author = {Emmanuel Clouet and Baptiste Bienvenu and Lucile Dezerald and David Rodney},
     title = {Screw dislocations in {BCC} transition metals: from \protect\emph{ab initio} modeling to yield criterion},
     journal = {C. R. Phys.},
     pages = {83--116},
     publisher = {Acad\'emie des sciences, Paris},
     volume = {22},
     number = {S3},
     year = {2021},
     doi = {10.5802/crphys.75}
}

@Article{nanoconfined_water_PD_Kapil2022,
author={Kapil, Venkat
and Schran, Christoph
and Zen, Andrea
and Chen, Ji
and Pickard, Chris J.
and Michaelides, Angelos},
title={The first-principles phase diagram of monolayer nanoconfined water},
journal={Nature},
year={2022},
month={Sep},
day={01},
volume={609},
number={7927},
pages={512-516},
issn={1476-4687},
doi={10.1038/s41586-022-05036-x},
url={https://doi.org/10.1038/s41586-022-05036-x}
}

@article{Gilbert_2008,
doi = {10.1088/0953-8984/20/34/345214},
url = {https://dx.doi.org/10.1088/0953-8984/20/34/345214},
year = {2008},
month = {aug},
publisher = {},
volume = {20},
number = {34},
pages = {345214},
author = {Gilbert, M R and Dudarev, S L and Derlet, P M and Pettifor, D G},
title = {Structure and metastability of mesoscopic vacancy and interstitial loop defects in iron and
tungsten},
journal = {J. Phys.: Condens. Matter}
}

@article{Mason_2017,
doi = {10.1088/1361-648X/aa9776},
url = {https://dx.doi.org/10.1088/1361-648X/aa9776},
year = {2017},
month = {nov},
publisher = {IOP Publishing},
volume = {29},
number = {50},
pages = {505501},
author = {Mason, D R and Nguyen-Manh, D and Becquart, C S},
title = {An empirical potential for simulating vacancy clusters in tungsten},
journal = {J. Phys.: Condens. Matter}
}

@article{SNAP_THOMPSON2015316,
title = {Spectral neighbor analysis method for automated generation of quantum-accurate interatomic potentials},
journal = {J. Comput. Phys.},
volume = {285},
pages = {316-330},
year = {2015},
issn = {0021-9991},
doi = {https://doi.org/10.1016/j.jcp.2014.12.018},
url = {https://www.sciencedirect.com/science/article/pii/S0021999114008353},
author = {A.P. Thompson and L.P. Swiler and C.R. Trott and S.M. Foiles and G.J. Tucker}
}

@Article{nanopart_surf_ener_relation_Yoko2018,
author={Yoko, Akira
and Umezawa, Naoto
and Ohno, Takahisa
and Oshima, Yoshito},
title={Impact of Surface Energy on the Formation of Composite Metal Oxide Nanoparticles},
journal={J. Phys. Chem. C},
year={2018},
month={Oct},
day={25},
publisher={American Chemical Society},
volume={122},
number={42},
pages={24350-24358},
issn={1932-7447},
doi={10.1021/acs.jpcc.8b06149},
url={https://doi.org/10.1021/acs.jpcc.8b06149}
}

@misc{amstools,
  note = {To access the \texttt{amstools}, contact the lead developer Yury Lysogorskiy (yury.lysogorskiy@icams.rub.de)}
}

@Inbook{Clouet2020-ab-initio,
author="Clouet, Emmanuel",
editor="Andreoni, Wanda
and Yip, Sidney",
title="Ab Initio Models of Dislocations",
bookTitle="Handbook of Materials Modeling: Methods: Theory and Modeling",
year="2020",
publisher="Springer International Publishing",
address="Cham",
pages="1503--1524",
isbn="978-3-319-44677-6",
doi="10.1007/978-3-319-44677-6_22",
url="https://doi.org/10.1007/978-3-319-44677-6_22"
}

@article{PhysRevLett.102.055502-Clouet-2009,
  title = {Dislocation Core Energies and Core Fields from First Principles},
  author = {Clouet, Emmanuel and Ventelon, Lisa and Willaime, F.},
  journal = {Phys. Rev. Lett.},
  volume = {102},
  issue = {5},
  pages = {055502},
  numpages = {4},
  year = {2009},
  month = {Feb},
  publisher = {American Physical Society},
  doi = {10.1103/PhysRevLett.102.055502},
  url = {https://link.aps.org/doi/10.1103/PhysRevLett.102.055502}
}

@article{ventelon2007corequadr,
  title={{Core structure and Peierls potential of screw dislocations in $\alpha$-Fe from first principles: cluster versus dipole approaches}},
  author={Ventelon, Lisa and Willaime, F},
  journal={J. Comp.-Aid. Mater. Des.},
  volume={14},
  number={Suppl 1},
  pages={85--94},
  year={2007},
  publisher={Springer},
  url={https://link.springer.com/article/10.1007/s10820-007-9064-y}
}

@article{Dezerald-dos-in-core,
  title = {Ab initio modeling of the two-dimensional energy landscape of screw dislocations in bcc transition metals},
  author = {Dezerald, L. and Ventelon, Lisa and Clouet, E. and Denoual, C. and Rodney, D. and Willaime, F.},
  journal = {Phys. Rev. B},
  volume = {89},
  issue = {2},
  pages = {024104},
  numpages = {13},
  year = {2014},
  month = {Jan},
  publisher = {American Physical Society},
  doi = {10.1103/PhysRevB.89.024104},
  url = {https://link.aps.org/doi/10.1103/PhysRevB.89.024104}
}

@article{self_diff_2_PhysRevB.80.144111,
  title = {{Molecular dynamics study of self-diffusion in bcc Fe}},
  author = {Mendelev, Mikhail I. and Mishin, Yuri},
  journal = {Phys. Rev. B},
  volume = {80},
  issue = {14},
  pages = {144111},
  numpages = {9},
  year = {2009},
  month = {Oct},
  publisher = {American Physical Society},
  doi = {10.1103/PhysRevB.80.144111},
  url = {https://link.aps.org/doi/10.1103/PhysRevB.80.144111}
}

@article{self_diff_1_PhysRevB.50.5928,
  title = {{Properties of monovacancies and self-interstitials in bcc Na: An ab initio pseudopotential study}},
  author = {Breier, U. and Frank, W. and Els\"asser, C. and F\"ahnle, M. and Seeger, A.},
  journal = {Phys. Rev. B},
  volume = {50},
  issue = {9},
  pages = {5928--5936},
  numpages = {0},
  year = {1994},
  month = {Sep},
  publisher = {American Physical Society},
  doi = {10.1103/PhysRevB.50.5928},
  url = {https://link.aps.org/doi/10.1103/PhysRevB.50.5928}
}

@article{exp_el_consts_simmons1965single,
  title={Single crystal elastic constants and calculated aggregate properties},
  author={Simmons, Gene},
  journal={J. Grad. Res. Ctr.},
  volume={34},
  number={1},
  pages={1},
  year={1965},
  url={https://scholar.smu.edu/journal_grc/vol34/iss1/1/}
}

@book{anderson2005fracture,
  title={Fracture mechanics: fundamentals and applications},
  author={Anderson, Ted L and Anderson, Ted L},
  year={2005},
  publisher={CRC press}
}

@book{anderson2017theory,
  title={Theory of dislocations},
  author={Anderson, Peter M and Hirth, John P and Lothe, Jens},
  year={2017},
  publisher={Cambridge University Press},
  address="Cambridge"
}

@article{el_const_mech_stab_Born_1940, title={On the stability of crystal lattices. I}, volume={36}, DOI={10.1017/S0305004100017138}, number={2}, journal={Math. Proc. Cambridge Philos. Soc.}, author={Born, Max}, year={1940}, pages={160–172}, url={https://www.cambridge.org/core/journals/mathematical-proceedings-of-the-cambridge-philosophical-society/article/abs/on-the-stability-of-crystal-lattices-i/E5FB4D987FC3FE1FF02DECDE41D550F1}}

@article{phonons_el_consts_RevModPhys.84.945,
  title = {Lattice instabilities in metallic elements},
  author = {Grimvall, G\"oran and Magyari-K\"ope, Blanka and Ozoli\ifmmode \mbox{\c{n}}\else \c{n}\fi{}\ifmmode \check{s}\else \v{s}\fi{}, Vidvuds and Persson, Kristin A.},
  journal = {Rev. Mod. Phys.},
  volume = {84},
  issue = {2},
  pages = {945--986},
  numpages = {0},
  year = {2012},
  month = {Jun},
  publisher = {American Physical Society},
  doi = {10.1103/RevModPhys.84.945},
  url = {https://link.aps.org/doi/10.1103/RevModPhys.84.945}
}

@article{Dudarev_vacancy_PhysRevMaterials.3.063601,
  title = {Effect of stress on vacancy formation and migration in body-centered-cubic metals},
  author = {Ma, Pui-Wai and Dudarev, S. L.},
  journal = {Phys. Rev. Mater.},
  volume = {3},
  issue = {6},
  pages = {063601},
  numpages = {12},
  year = {2019},
  month = {Jun},
  publisher = {American Physical Society},
  doi = {10.1103/PhysRevMaterials.3.063601},
  url = {https://link.aps.org/doi/10.1103/PhysRevMaterials.3.063601}
}

@Article{MatBench_Riebesell2025,
author={Riebesell, Janosh
and Goodall, Rhys E. A.
and Benner, Philipp
and Chiang, Yuan
and Deng, Bowen
and Ceder, Gerbrand
and Asta, Mark
and Lee, Alpha A.
and Jain, Anubhav
and Persson, Kristin A.},
title={A framework to evaluate machine learning crystal stability predictions},
journal={Nat. Mach. Intell.},
year={2025},
month={Jun},
day={01},
volume={7},
number={6},
pages={836-847},
issn={2522-5839},
doi={10.1038/s42256-025-01055-1},
url={https://doi.org/10.1038/s42256-025-01055-1}
}

@article{GRACE_PhysRevX.14.021036,
  title = {Graph Atomic Cluster Expansion for Semilocal Interactions beyond Equivariant Message Passing},
  author = {Bochkarev, Anton and Lysogorskiy, Yury and Drautz, Ralf},
  journal = {Phys. Rev. X},
  volume = {14},
  issue = {2},
  pages = {021036},
  numpages = {28},
  year = {2024},
  month = {Jun},
  publisher = {American Physical Society},
  doi = {10.1103/PhysRevX.14.021036},
  url = {https://link.aps.org/doi/10.1103/PhysRevX.14.021036}
}

@phdthesis{Rui_Wang_phd,
  author={Wang, R.},
  year={2023},
  school={City University of Hong Kong},
  url={https://scholars.cityu.edu.hk/en/studentTheses/developing-extended-modified-embedded-atom-method-potentials-for-},
  title={{Developing Extended Modified Embedded-Atom Method Potentials for Atomistic Modelling on Plasticity and Fracture Behaviours of Metals}}
}

@Article{MTP-2_Jung2023,
author={Jung, Jong Hyun
and Srinivasan, Prashanth
and Forslund, Axel
and Grabowski, Blazej},
title={High-accuracy thermodynamic properties to the melting point from ab initio calculations aided by machine-learning potentials},
journal={npj Comput. Mater.},
year={2023},
month={Jan},
day={10},
volume={9},
number={1},
pages={3},
issn={2057-3960},
doi={10.1038/s41524-022-00956-8},
url={https://doi.org/10.1038/s41524-022-00956-8}
}

@Article{MTP-1_Yin2021,
author={Yin, Sheng
and Zuo, Yunxing
and Abu-Odeh, Anas
and Zheng, Hui
and Li, Xiang-Guo
and Ding, Jun
and Ong, Shyue Ping
and Asta, Mark
and Ritchie, Robert O.},
title={Atomistic simulations of dislocation mobility in refractory high-entropy alloys and the effect of chemical short-range order},
journal={Nat. Commun.},
year={2021},
month={Aug},
day={11},
volume={12},
number={1},
pages={4873},
issn={2041-1723},
doi={10.1038/s41467-021-25134-0},
url={https://doi.org/10.1038/s41467-021-25134-0}
}

@article{EAM_PhysRevB.81.144119,
  title = {Force-matched embedded-atom method potential for niobium},
  author = {Fellinger, Michael R. and Park, Hyoungki and Wilkins, John W.},
  journal = {Phys. Rev. B},
  volume = {81},
  issue = {14},
  pages = {144119},
  numpages = {15},
  year = {2010},
  month = {Apr},
  publisher = {American Physical Society},
  doi = {10.1103/PhysRevB.81.144119},
  url = {https://link.aps.org/doi/10.1103/PhysRevB.81.144119}
}

@article{Kunzmann_PhysRevMaterials.8.033603,
  title = {{Ab initio study of transition paths between (meta)stable phases of Nb and Ta-substituted Nb}},
  author = {Kunzmann, Susanne and Hammerschmidt, Thomas and Schierning, Gabi and Gr\"unebohm, Anna},
  journal = {Phys. Rev. Mater.},
  volume = {8},
  issue = {3},
  pages = {033603},
  numpages = {8},
  year = {2024},
  month = {Mar},
  publisher = {American Physical Society},
  doi = {10.1103/PhysRevMaterials.8.033603},
  url = {https://link.aps.org/doi/10.1103/PhysRevMaterials.8.033603}
}

@article{GAP_Nb_PhysRevMaterials.4.093802,
  title = {Gaussian approximation potentials for body-centered-cubic transition metals},
  author = {Byggm\"astar, J. and Nordlund, K. and Djurabekova, F.},
  journal = {Phys. Rev. Mater.},
  volume = {4},
  issue = {9},
  pages = {093802},
  numpages = {11},
  year = {2020},
  month = {Sep},
  publisher = {American Physical Society},
  doi = {10.1103/PhysRevMaterials.4.093802},
  url = {https://link.aps.org/doi/10.1103/PhysRevMaterials.4.093802}
}

@article{pacemaker-Lysogorskiy2021,
  author = {Lysogorskiy, Y. and van der Oord, C. and Bochkarev, A. and Menon, S. and Rinaldi, M. and Hammerschmidt, T. and Mrovec, M. and Thompson, A. and Csányi, G. and Ortner, C. and Drautz, R.},
  title = {{Performant implementation of the atomic cluster expansion (PACE) and application to copper and silicon}},
  journal = {npj Comput. Mater.},
  volume = {7},
  pages = {97},
  year = {2021},
  doi = {10.1038/s41524-021-00559-9}
}

@article{zhang-maresca-2023GAPs,
title = {{Efficiency, accuracy, and transferability of machine learning potentials: Application to dislocations and cracks in iron}},
journal = {Acta Mater.},
volume = {270},
pages = {119788},
year = {2024},
issn = {1359-6454},
doi = {https://doi.org/10.1016/j.actamat.2024.119788},
url = {https://www.sciencedirect.com/science/article/pii/S135964542400140X},
author = {Lei Zhang and Gábor Csányi and Erik {van der Giessen} and Francesco Maresca}
}

@article{tuckerman_cendagorta2021enhanced,
  title={Enhanced sampling path integral methods using neural network potential energy surfaces with application to diffusion in hydrogen hydrates},
  author={Cendagorta, Joseph R and Shen, Hengyuan and Ba{\v{c}}i{\'c}, Zlatko and Tuckerman, Mark E},
  journal={Adv. Theory Simul.},
  volume={4},
  number={4},
  pages={2000258},
  year={2021},
  publisher={Wiley Online Library},
  url={https://advanced.onlinelibrary.wiley.com/doi/full/10.1002/adts.202000258}
}

@article{minaam_ngoipala2025hydride,
  title={Hydride-Induced Reconstruction of Pd Electrode Surfaces: A Combined Computational and Experimental Study},
  author={Ngoipala, Apinya and Schott, Christian and Briega-Martos, Valentin and Qamar, Minaam and Mrovec, Matous and Javan Nikkhah, Sousa and Schmidt, Thorsten O and Deville, Lewin and Capogrosso, Andrea and Moumaneix, Lilian and others},
  journal={Adv. Mater.},
  volume={37},
  number={4},
  pages={2410951},
  year={2025},
  publisher={Wiley Online Library},
  url={https://advanced.onlinelibrary.wiley.com/doi/full/10.1002/adma.202410951}
}

@article{
freitas_doi:10.1073/pnas.2322962121,
author = {Killian Sheriff  and Yifan Cao  and Tess Smidt  and Rodrigo Freitas },
title = {Quantifying chemical short-range order in metallic alloys},
journal = {Proc. Natl Acad. Sci. USA},
volume = {121},
number = {25},
pages = {e2322962121},
year = {2024},
doi = {10.1073/pnas.2322962121},
URL = {https://www.pnas.org/doi/abs/10.1073/pnas.2322962121}}

@article{
cheng_water_doi:10.1073/pnas.1815117116,
author = {Bingqing Cheng  and Edgar A. Engel  and Jörg Behler  and Christoph Dellago  and Michele Ceriotti },
title = {Ab initio thermodynamics of liquid and solid water},
journal = {Proc. Natl Acad. Sci. USA},
volume = {116},
number = {4},
pages = {1110-1115},
year = {2019},
doi = {10.1073/pnas.1815117116},
URL = {https://www.pnas.org/doi/abs/10.1073/pnas.1815117116}
}

@Article{hydrogen_Cheng2020,
author={Cheng, Bingqing
and Mazzola, Guglielmo
and Pickard, Chris J.
and Ceriotti, Michele},
title={Evidence for supercritical behaviour of high-pressure liquid hydrogen},
journal={Nature},
year={2020},
month={Sep},
day={01},
volume={585},
number={7824},
pages={217-220},
issn={1476-4687},
doi={10.1038/s41586-020-2677-y},
url={https://doi.org/10.1038/s41586-020-2677-y}
}

@Article{Kozinsky_potential_Batzner2022,
author={Batzner, Simon
and Musaelian, Albert
and Sun, Lixin
and Geiger, Mario
and Mailoa, Jonathan P.
and Kornbluth, Mordechai
and Molinari, Nicola
and Smidt, Tess E.
and Kozinsky, Boris},
title={E(3)-equivariant graph neural networks for data-efficient and accurate interatomic potentials},
journal={Nat. Commun.},
year={2022},
month={May},
day={04},
volume={13},
number={1},
pages={2453},
doi={10.1038/s41467-022-29939-5},
url={https://doi.org/10.1038/s41467-022-29939-5}
}

@inproceedings{MACE_NEURIPS2022_4a36c3c5,
 author = {Batatia, Ilyes and Kovacs, David P and Simm, Gregor and Ortner, Christoph and Csanyi, Gabor},
 booktitle = {Adv. Neural Inf. Process. Syst.},
 editor = {S. Koyejo and S. Mohamed and A. Agarwal and D. Belgrave and K. Cho and A. Oh},
 pages = {11423--11436},
 publisher = {Curran Associates, Inc.},
 title = {MACE: Higher Order Equivariant Message Passing Neural Networks for Fast and Accurate Force Fields},
 url = {https://proceedings.neurips.cc/paper_files/paper/2022/file/4a36c3c51af11ed9f34615b81edb5bbc-Paper-Conference.pdf},
 volume = {35},
 year = {2022}
}

@article{Thygesen_2D_van_der_Waals,
  title = {{Dispersion-corrected machine learning potentials for 2D van der Waals materials}},
  author = {Sauer, Mikkel Ohm and Lyngby, Peder Meisner and Thygesen, Kristian Sommer},
  journal = {Phys. Rev. Mater.},
  volume = {9},
  issue = {7},
  pages = {074007},
  numpages = {11},
  year = {2025},
  month = {Jul},
  publisher = {American Physical Society},
  doi = {10.1103/cl8c-8f1f},
  url = {https://link.aps.org/doi/10.1103/cl8c-8f1f}
}

@article{dft_c44_nb_underestim_10.1063/1.5136052,
    author = {Wang, Yi X. and Geng, Hua Y. and Wu, Q. and Chen, Xiang R.},
    title = {Orbital localization error of density functional theory in shear properties of vanadium and niobium},
    journal = {J. Chem. Phys.},
    volume = {152},
    number = {2},
    pages = {024118},
    year = {2020},
    month = {01},
    issn = {0021-9606},
    doi = {10.1063/1.5136052},
    url = {https://doi.org/10.1063/1.5136052},
}

@article{Deringer_Pickard_boron_PhysRevLett.120.156001,
  title = {Data-Driven Learning of Total and Local Energies in Elemental Boron},
  author = {Deringer, Volker L. and Pickard, Chris J. and Cs\'anyi, G\'abor},
  journal = {Phys. Rev. Lett.},
  volume = {120},
  issue = {15},
  pages = {156001},
  numpages = {5},
  year = {2018},
  month = {Apr},
  publisher = {American Physical Society},
  doi = {10.1103/PhysRevLett.120.156001},
  url = {https://link.aps.org/doi/10.1103/PhysRevLett.120.156001}
}

@article{surface_PhysRevLett.125.206101,
  title = {{${\mathrm{IrO}}_{2}$ Surface Complexions Identified through Machine Learning and Surface Investigations}},
  author = {Timmermann, Jakob and Kraushofer, Florian and Resch, Nikolaus and Li, Peigang and Wang, Yu and Mao, Zhiqiang and Riva, Michele and Lee, Yonghyuk and Staacke, Carsten and Schmid, Michael and Scheurer, Christoph and Parkinson, Gareth S. and Diebold, Ulrike and Reuter, Karsten},
  journal = {Phys. Rev. Lett.},
  volume = {125},
  issue = {20},
  pages = {206101},
  numpages = {6},
  year = {2020},
  month = {Nov},
  publisher = {American Physical Society},
  doi = {10.1103/PhysRevLett.125.206101},
  url = {https://link.aps.org/doi/10.1103/PhysRevLett.125.206101}
}

@article{alchemical_PhysRevMaterials.7.045802,
  title = {Modeling high-entropy transition metal alloys with alchemical compression},
  author = {Lopanitsyna, Nataliya and Fraux, Guillaume and Springer, Maximilian A. and De, Sandip and Ceriotti, Michele},
  journal = {Phys. Rev. Mater.},
  volume = {7},
  issue = {4},
  pages = {045802},
  numpages = {15},
  year = {2023},
  month = {Apr},
  publisher = {American Physical Society},
  doi = {10.1103/PhysRevMaterials.7.045802},
  url = {https://link.aps.org/doi/10.1103/PhysRevMaterials.7.045802}
}

@article{Oganov_MLIPs_for_CSP_PhysRevB.99.064114,
  title = {Accelerating crystal structure prediction by machine-learning interatomic potentials with active learning},
  author = {Podryabinkin, Evgeny V. and Tikhonov, Evgeny V. and Shapeev, Alexander V. and Oganov, Artem R.},
  journal = {Phys. Rev. B},
  volume = {99},
  issue = {6},
  pages = {064114},
  numpages = {7},
  year = {2019},
  month = {Feb},
  publisher = {American Physical Society},
  doi = {10.1103/PhysRevB.99.064114},
  url = {https://link.aps.org/doi/10.1103/PhysRevB.99.064114}
}

@Article{deringer_device_ACE_zhou2025fullcycledevicescalesimulationsmemory,
author={Zhou, Yuxing
and Thomas du Toit, Daniel F.
and Elliott, Stephen R.
and Zhang, Wei
and Deringer, Volker L.},
title={Full-cycle device-scale simulations of memory materials with a tailored atomic-cluster-expansion potential},
journal={Nat. Commun.},
year={2025},
month={Sep},
day={30},
volume={16},
number={1},
pages={8688},
issn={2041-1723},
doi={10.1038/s41467-025-63732-4},
url={https://doi.org/10.1038/s41467-025-63732-4}
}

@Article{Deringer_Device_GAP_Zhou2023,
author={Zhou, Yuxing
and Zhang, Wei
and Ma, En
and Deringer, Volker L.},
title={Device-scale atomistic modelling of phase-change memory materials},
journal={Nat. Electron.},
year={2023},
month={Oct},
day={01},
volume={6},
number={10},
pages={746-754},
issn={2520-1131},
doi={10.1038/s41928-023-01030-x},
url={https://doi.org/10.1038/s41928-023-01030-x}
}

@Article{Deringer_Nature_2021,
author={Deringer, Volker L.
and Bernstein, Noam
and Cs{\'a}nyi, G{\'a}bor
and Ben Mahmoud, Chiheb
and Ceriotti, Michele
and Wilson, Mark
and Drabold, David A.
and Elliott, Stephen R.},
title={Origins of structural and electronic transitions in disordered silicon},
journal={Nature},
year={2021},
month={Jan},
day={01},
volume={589},
number={7840},
pages={59-64},
issn={1476-4687},
doi={10.1038/s41586-020-03072-z},
url={https://doi.org/10.1038/s41586-020-03072-z}
}

@Article{Meta_Lan2023,
author={Lan, Janice
and Palizhati, Aini
and Shuaibi, Muhammed
and Wood, Brandon M.
and Wander, Brook
and Das, Abhishek
and Uyttendaele, Matt
and Zitnick, C. Lawrence
and Ulissi, Zachary W.},
title={{AdsorbML: a leap in efficiency for adsorption energy calculations using generalizable machine learning potentials}},
journal={npj Comput. Mater.},
year={2023},
month={Sep},
day={22},
volume={9},
number={1},
pages={172},
issn={2057-3960},
doi={10.1038/s41524-023-01121-5},
url={https://doi.org/10.1038/s41524-023-01121-5}
}

@Article{W_diffusion_Grabowski_Zhang2025,
author={Zhang, Xi
and Divinski, Sergiy V.
and Grabowski, Blazej},
title={Ab initio machine-learning unveils strong anharmonicity in non-Arrhenius self-diffusion of tungsten},
journal={Nat. Commun.},
year={2025},
month={Jan},
day={04},
volume={16},
number={1},
pages={394},
issn={2041-1723},
doi={10.1038/s41467-024-55759-w},
url={https://doi.org/10.1038/s41467-024-55759-w}
}

@article{universal_3_Shuang_2025,
doi = {10.1088/2632-2153/adea2d},
url = {https://dx.doi.org/10.1088/2632-2153/adea2d},
year = {2025},
month = {jul},
publisher = {IOP Publishing},
volume = {6},
number = {3},
pages = {030501},
author = {Shuang, Fei and Wei, Zixiong and Liu, Kai and Gao, Wei and Dey, Poulumi},
title = {{Universal machine learning interatomic potentials poised to supplant DFT in modeling general defects in metals and random alloys}},
journal = {Mach. Learn. Sci. Tech.}
}

\end{document}